\documentclass[final,5p,times,singlecolumn]{elsarticle}
\usepackage{amsmath,latexsym,mathtools}
\usepackage{soul}
\usepackage[x11names,dvipsnames,table]{xcolor} 

\usepackage{multicol}
\usepackage{dcolumn}
\usepackage{subcaption}
\usepackage{caption}

\usepackage[pscoord]{eso-pic}
\newcommand{\placetextbox}[3]{
\setbox0=\hbox{#3}
\AddToShipoutPictureFG*{
\put(\LenToUnit{#1\paperwidth},\LenToUnit{#2\paperheight}){\vtop{{\null}\makebox[0pt][c]{#3}}}%
}%
}%

\usepackage{makecell}
\usepackage{booktabs} 
\usepackage[normalem]{ulem}

\usepackage{etoolbox}
\usepackage{amsmath,latexsym,mathtools}
\usepackage{comment}
\usepackage{footnpag}
\usepackage{float} 
\usepackage{xcolor}    
\usepackage[multiple]{footnotehyper}
\usepackage{chngcntr}
\journal{Physics of the Dark Universe}
\counterwithout{footnote}{section} 

\usepackage{slashed}
\usepackage{braket}
\usepackage{amssymb}
\usepackage{amsfonts}
\usepackage{float}
\usepackage[utf8]{inputenc}
\usepackage{amsmath}
\usepackage{physics}
\usepackage{graphicx} 
\usepackage{dcolumn} 
\usepackage{bm} 

\usepackage{subcaption}
\usepackage{changepage}
\usepackage{xcolor}
\usepackage{float} 
\journal{Physics of the Dark Universe}

\usepackage[colorlinks]{hyperref}
\usepackage[thinc]{esdiff}

\AtBeginDocument{\hypersetup{
	colorlinks=true,
	linkcolor=black!40!blue,
	citecolor=black!40!blue,
	filecolor=black,
	urlcolor=black!45!blue,
}}

\usepackage{comment}

\begin{document}

\title{Cosmological perturbations of TDiff fields}

\author[1]{Antonio L. Maroto}
\ead{maroto@ucm.es}

\author[1]{Prado Martín-Moruno}
\ead{pradomm@ucm.es}

\author[1]{Diego Tessainer \corref{cor1}}
\ead{dtessain@ucm.es}

\cortext[cor1]{Corresponding author}

\affiliation[1]{organization={Departamento de Física Teórica and Instituto de Física de Partículas y del Cosmos (IPARCOS-UCM)}, addressline={\\Universidad Complutense de Madrid},postcode={28040}, city={Madrid}, country={Spain.}}

\begin{abstract}
We study scalar field theories that break diffeomorphism invariance down to the subgroup of transverse diffeomorphisms through the matter sector in cosmological backgrounds. We focus on single- and multi-field models and develop the corresponding cosmological perturbation theory. We analyze the different  contributions to the pressure perturbation, discussing the adiabaticity and the effects in the perturbation coefficients of the interactions that arise in the multi-field case as a consequence of the symmetry breaking. We also consider the stability of the perturbations in terms of the effective speed of sound and present particular models that could be of phenomenological interest. 
\end{abstract}


\maketitle

\placetextbox{0.85}{0.95}{IPARCOS-UCM-26-022}
\section{Introduction}
Over the last few years, theories invariant under transverse diffeomorphisms (TDiff) have gained popularity. Even if diffeomorphism (Diff) invariance (which translates into physical invariance under general coordinate transformations) is the fundamental symmetry of General Relativity (GR), certain  observational and theoretical issues  may suggest introducing modifications of the theory at cosmological scales \cite{Clifton:2011jh}. In particular, the unknown nature of the dark sector \cite{Copeland:2006wr} or the observational tensions of the cosmological parameters \cite{SupernovaSearchTeam:1998fmf,Riess:2019cxk,DiValentino:2020zio} are,  to date, the most pressing problems.

TDiff invariant models draw inspiration from unimodular gravity \cite{einsteinUG,PhysRevD.40.1048,Carballo-Rubio_2022}, in which the metric determinant is a fixed non-dynamical field ($g=1$) and Diff invariance is broken down to transverse diffeomorphisms and Weyl rescalings, having been shown to provide a solution to the vacuum-energy problem \cite{Ellis:2010uc}. TDiff gravity beyond unimodular gravity has been studied in \cite{Toni1,Toni2:2024vqk}, showing that an additional scalar graviton mode propagates in such theories. On the other hand, theories breaking Diff invariance through the matter sector have also been investigated in \cite{Maroto_Primigenio,Dario1,Dario_cov,DAWEI,Tessainer1,Maroto:2025bxy,JavierdeCruzPerez:2025ytd,Alfredo1:2024mkx,Alfredo2:2024roe,BeltranJimenez:2025pho,Maroto:2025wtg}. More specifically, \cite{Maroto_Primigenio,Dario1,DAWEI,JavierdeCruzPerez:2025ytd} focus on the study of single-field TDiff theories at the background level, with \cite{DAWEI} and \cite{JavierdeCruzPerez:2025ytd} introducing more phenomenological applications concerning the dark sector. In \cite{Dario_cov}, a way to describe TDiff invariant theories is introduced through covariantization, and in \cite{Maroto:2025wtg} the consequences of breaking Diff invariance in the inflaton sector are explored. Ref. \cite{BeltranJimenez:2025pho} focuses on the study of equivalences between TDiff invariant theories and $k$-essence or mimetic theories, while \cite{Alfredo1:2024mkx,Alfredo2:2024roe} focus on TDiff models for single abelian gauge fields and their implications for cosmic magnetic evolution. When it comes to multi-field TDiff theories, the analysis  performed in \cite{Tessainer1,Maroto:2025bxy} for certain models shows that the symmetry breaking results in an effective interaction taking place between the fields and alters their dynamics and time evolution, which is of particular interest when it comes to the dark sector. 

However, despite the previous results, to date, TDiff theories have been mostly studied at the background level, and a detailed analysis of cosmological perturbations is still lacking. Thus, in this work we will focus on establishing the cosmological perturbation theory and applying it to single-field and multi-field models. We will investigate the effects of the interactions when it comes to the effective speed of sound and analyze the different contributions to the pressure perturbations, discussing the stability of the models. 

The paper is organized as follows. In section \ref{sec2}, we will present the TDiff formalism and the covariantized approach, and review the main results obtained for single- and multi-field theories at the background level. Section \ref{sec3} is devoted to the development of cosmological perturbation theory in the single-field case, where the pressure perturbation is computed and the kinetic and potential domination regimes are also discussed. Section \ref{sec4} is dedicated to the application of the perturbative formalism in the context of multi-field TDiff theories. The different contributions to the pressure perturbation will be calculated, and the adiabaticity and the effects of the interactions will be discussed. We will also analyze the stability of the models in terms of the speed of sound and compare TDiff models with the analogous Diff invariant ones. In section \ref{sec5}, we will briefly present particular models with non-negative speed of sound that could be of relevant interest from a phenomenological point of view. Lastly, in section \ref{conclusions} we will present the main conclusions.  

\section{TDiff action and the covariantized approach}\label{sec2}
In this section we will first introduce the concept of transverse diffeomorphisms \cite{Maroto_Primigenio} and present the TDiff invariant action for a matter sector consisting of a scalar field $\phi$ and review the results obtained for single TDiff scalar field theories when working using the more convenient covariantized approach (see \cite{Dario_cov}). Afterwards, in section \ref{sec22} we will present and discuss the most relevant aspects of multi-field TDiff theories involving several scalar fields at the background level \cite{Tessainer1,Maroto:2025bxy}. 
\subsection{Single-field TDiff theories}
Let us consider a general (infinitesimal) coordinate transformation given by a vector field $\xi^\mu(x)$, i.e., $x^\mu\mapsto\hat{x}^\mu=x^\mu+\xi^\mu(x)$. The transformation of the metric tensor $g_{\mu\nu}(x)$ will be given by the Lie derivative $\mathfrak{L}_\xi(g_{\mu\nu})$ according to:
\begin{equation}
    \delta g_{\mu\nu}=\mathfrak{L}_\xi(g_{\mu\nu})=-\nabla_\nu\xi_\mu-\nabla_\mu\xi_\nu.
    \label{delta_g_mu_nu}
\end{equation}
Consequently, the metric determinant $g\equiv|\mathrm{det}(g_{\mu\nu})|$ will transform as
\begin{equation}
    \delta g=gg^{\mu\nu}\delta g_{\mu\nu}=-2g\nabla _\mu\xi^\mu.
    \label{delta_g}
\end{equation}
Let us now introduce our action, which will involve a gravitational and a matter contribution. Since Diff invariance will only be broken through the matter sector, the gravitational term will simply be the usual Einstein-Hilbert action. Therefore, our action will read
\begin{equation}
    S=S_\mathrm{EH}[g_{\mu\nu}]+S_\mathrm{mat}[g_{\mu\nu},\phi],
    \label{S_tot}
\end{equation}
where $S_\mathrm{mat}$ is the matter action associated with the scalar field,
\begin{equation}
    S_\mathrm{mat}[g_{\mu\nu},\phi]=\int \dd^4 xf(g)\mathcal{L}(g_{\mu\nu}(x),\phi(x),\partial_\mu\phi(x)),
    \label{S_mat}
\end{equation}
with $f(g)$ an arbitrary coupling function of the metric determinant $g$; and $S_\mathrm{EH}$ is the Einstein-Hilbert action, given by
\begin{equation}
    S_\mathrm{EH}[g_{\mu\nu}]=-\frac{1}{16\pi G}\int \dd^4x\sqrt{g}R.
    \label{S_EH}
\end{equation}
The variation of the matter action under the coordinate transformation given by $\xi^\mu$ will thus read 
\begin{equation}
    \delta_\xi S_\mathrm{mat}=\int\dd^4 x\partial_\mu\xi^\mu[f(g)-2gf'(g)]\mathcal{L},
    \label{delta_S_mat}
\end{equation}
where from now on the prime indices will denote derivatives with respect to the argument of the respective functions. Looking at \eqref{delta_S_mat}, we can observe that the action will remain invariant under these transformations if $f(g)-2gf'(g)=0$, which, as expected, is equivalent to having Diff invariance ($f(g)\propto \sqrt{g}$). However, if the vector field driving the transformation is divergenceless, i.e. $\partial_\mu\xi^\mu=0$, the action will be invariant for arbitrary $f(g)$. This is called the transversality condition and it is required for the coordinate transformation to be a transverse diffeomorphism. Notice that under transverse diffeomorphisms, the metric determinant $g$ transforms as a true scalar \cite{Maroto_Primigenio}. 

The generic form of the TDiff matter action for a scalar field $\phi$ with a canonical kinetic term will read:
\begin{equation}
    S_\mathrm{mat}=\int\dd^4 x\left[\frac{1}{2}f_K(g)g^{\mu\nu}\partial_\mu\phi\partial_\nu\phi-f_V(g)V(\phi)\right],
    \label{S_scalar_single}
\end{equation}
where $f_K(g)$ and $f_V(g)$ denote the respective coupling functions of the kinetic and potential terms, and they are chosen to be positive-valued to avoid the appearance of instabilities.

After having presented the general TDiff action for a scalar field, we will now introduce the covariantized approach \cite{Dario_cov}, which is much more convenient to work with at the perturbative level and, in fact, is the formalism we will use in the rest of the paper. 
The covariantized formalism presented in \cite{Dario_cov} follows Henneaux and Teitelboim \cite{Henneaux_Teitelboim} and allows us to write our theory in a covariantized way through the introduction of a vector field $A^\mu$, similarly to the Stueckelberg approach in gauge field theories \cite{STUECKELBERG1}. If we define $Y\equiv \nabla_\mu A^\mu$, the covariantized action will thus read \cite{Dario_cov}:
\begin{equation}
    S_\mathrm{mat}=\int \dd^4 x\sqrt{g}[H_K(Y)X-H_V(Y)V(\phi)],
    \label{S_1scalar_gen_cov}
\end{equation}
where 
\begin{align}
X\equiv \frac{1}{2}g^{\mu\nu}\partial_\mu\phi\partial_\nu\phi 
\end{align}
depicts the canonical kinetic term and where the $H_i$, $i=K,V$ are related to the $f_i$ via $H_i(Y)=Yf_i(Y^{-2})$. This action is equivalent to the TDiff action \eqref{S_scalar_single} in the coordinate  frame in which $Y= 1/\sqrt{g}$, which we shall call the TDiff frame \cite{Dario_cov}. Consequently, the EoM for the scalar field $\phi$ and the Stueckelberg field $A^\mu$ will respectively read
\begin{equation}
    \nabla_\nu[H_K(Y)\nabla^\mu\phi]+H_V(Y)V'(\phi)=0,
    \label{EoM_phi_gen}
\end{equation}
and
\begin{equation}
    \partial_\nu[H_K'(Y)X-H_V'(Y)V(\phi)]=0.
\end{equation}
which is equivalent to the existence of a constant of motion
\begin{align}
H_K'(Y)X-H_V'(Y)V(\phi)=C_g.    
\end{align}
On the other hand, the energy-momentum tensor (EMT) can be calculated by performing variations of the action with respect to the metric
\begin{equation}
    T_{\mu\nu}=\frac{2}{\sqrt{g}}\frac{\delta S_\mathrm{mat}}{\delta g^{\mu\nu}},
    \label{EMT_def}
\end{equation}
which, in fact, takes the perfect fluid form
\begin{equation}
    T_{\mu\nu}=(\rho+p)u_\mu u_\nu-pg_{\mu\nu},
    \label{EMT_perfect_fluid}
\end{equation}
where $u_\mu\equiv\partial_\mu\phi/\mathcal{N}$ is the four-velocity (if $u_\mu$ is a timelike vector) and $\mathcal{N}\equiv\sqrt{\partial_\mu\phi\partial^\mu\phi}$ is a normalization constant. $\rho$ and $p$ are the energy density and pressure, respectively, which read
\begin{eqnarray}
    \rho=X[H_K(Y)+YH_K'(Y)]+V(\phi)[H_V(Y)-YH_V'(Y)],
    \label{rho_perfect_fluid}\\
    p=X[H_K(Y)-YH_K'(Y)]-V(\phi)[H_V(Y)-YH_V'(Y)].
    \label{p_perfect_fluid}
\end{eqnarray}
The EMT will be automatically conserved under solutions of the field EoM since we have written the theory in a Diff invariant way using the covariantized approach. 

Lastly, it is worth mentioning that, since we are going to work in cosmological backgrounds, we will consider our background to be the spatially flat Friedmann-Lemaître-Robertson-Walker spacetime:
\begin{equation}
    \dd s^2=a(\eta)^2(\dd\eta^2-\dd\mathbf{x}^2),
    \label{flat_FLRW}
\end{equation}
where $\eta$ is the conformal time. We shall now discuss the different domination regimes.

\subsubsection{Potential domination regime}
Let us now focus on the potential domination regime. We will define this regime as the one in which the field varies slowly with respect to the other scales, i.e., $X\rightarrow0$. Thus, since the potential term is dominant in the action \eqref{S_1scalar_gen_cov}, $V(\phi)$ will be approximately constant:
\begin{equation}
    V(\phi)=\mathrm{const.}
    \label{EoM_phi_1pot}
\end{equation}
Consequently, the EoM for the Stueckelberg field $A^\mu$ reads, in this case, 
\begin{equation}
    H_V'(Y)V(\phi)=\mathrm{const.},
    \label{EoM_Amu_1pot}
\end{equation}
which implies that $Y=\text{const.}$ and thus $g=\text{const.}$ in the TDiff frame \footnote{Notice that there will be higher order corrections to this solution, but they will be small and act as slow-roll contributions.}.

\subsubsection{Kinetic domination regime}
If the kinetic term dominates the action of the scalar field, the EoM for the Stueckelberg field and for $\phi$ will respectively read
\begin{equation}
    H_K'(Y)X=\text{const.},
    \label{EoM_Amu_1kin}
\end{equation}
and 
\begin{equation}
    \nabla_\mu[H_K(Y)\partial^\mu\phi]=\nabla_\mu[H_K(Y)\mathcal{N}u^\mu]=0;
    \label{EoM_phi_1kin}
\end{equation}
where \eqref{EoM_phi_1kin} yields \cite{Dario_cov}:
\begin{equation}
    2X=\frac{C_\phi^2}{H_K(Y)^2a^6},
    \label{constr_1kin_step1}
\end{equation}
where $C_\phi$ is a constant. Thus, substituting this expression into \eqref{EoM_Amu_1kin} results in the following condition: 
\begin{equation}
    -\frac{H_K'(Y)}{H_K(Y)^2}=Ca^6,
    \label{constr_1kin}
\end{equation}
with $C$ a constant. Lastly, the expressions for the energy density and pressure of $\phi$ will read, under kinetic domination, 
\begin{align}
    \rho&=X[H_K(Y)+YH_K'(Y)],
    \label{rho1kin_cov}\\
    p&=w\rho,
    \label{p1kin_cov}
\end{align}
where 
\begin{equation}
    w=\frac{H_K(Y)-YH_K'(Y)}{H_K(Y)+YH_K'(Y)}
    \label{w_1kin_general}
\end{equation}
depicts the equation of state (EoS) parameter, which is a function of $Y$ and will generally evolve with the expansion of the universe, unlike in the Diff case, where just stiff behavior is possible for purely kinetically driven fields. 

\subsection{Multi-field TDiff theories}\label{sec22}
We will now present the particular models we will be working with in the rest of the work, which involve two free TDiff homogeneous scalar fields in different regimes. We will briefly recap their general cosmological consequences at the background level (see \cite{Tessainer1,Maroto:2025bxy} for a detailed review and analysis) and discuss the power-law coupling, which is very interesting from a physical point of view. The most general TDiff matter action involving two non-interacting scalar fields to the lowest order in derivatives reads, in the covariantized approach,
\begin{align}
    S_\mathrm{mat}&=\int \dd^4 x\sqrt{g}[H_{K_1}(Y)X_1-H_{V_1}(Y)V_1(\phi_1)\nonumber   \\
    &+H_{K_2}(Y)X_2-H_{V_2}(Y)V_2(\phi_2)].
    \label{S_mat_gen_2fields}
\end{align}
The most relevant cases of interest for these models were studied in \cite{Tessainer1,Maroto:2025bxy} at the background level. On the one hand, there is the shift-symmetric case \cite{Tessainer1}, in which both scalar fields are dominated by their kinetic term and, on the other hand, there is the mixed-regime case \cite{Maroto:2025bxy}, in which one of the fields exhibits shift-symmetry and is kinetically driven, while the other is dominated by its potential contribution for a sufficiently long period of time. Notice that the fields are not interacting directly since there is no interaction potential between them and the total EMT is simply
\begin{equation}
    T_{\mu\nu}=T_{\mu\nu}^{(1)}+T_{\mu\nu}^{(2)},
    \label{EMT_total}
\end{equation}
where 
\begin{equation}
    T_{\mu\nu}^{(i)}=(\rho_i+p_i)u_\mu u_\nu-p_ig_{\mu\nu},
    \label{EMT_individual}
\end{equation}
with both components sharing a common four-velocity for being homogeneous. However, even if the fields are not interacting directly, the Diff symmetry breaking (the presence of an extra field in the covariantized approach) results in an effective interaction taking place between the fluids as we will later see in further detail. In fact, as discussed previously, the total EMT must be conserved:
\begin{equation}
    \nabla_\mu T^{\mu\nu}=\nabla_\mu T^{(1)\mu\nu}+\nabla_\mu T^{(2)\mu\nu}=0,
\end{equation}
but nothing is imposed on the individual EMTs. In the flat FLRW background, we could parametrize the non-conservation writing the EMT conservation equations as
\begin{align}
    \dot{\rho}_1+3\frac{\dot{a}}{a}[\rho_1+p_1]=&Q,
    \label{cons1Q}\\
    \dot{\rho}_2+3\frac{\dot{a}}{a}[\rho_2+p_2]=&-Q,
    \label{cons2Q}
\end{align}
where $Q$ is the interacting kernel and where from now on the dot indices will denote derivatives with respect to conformal time $\eta$. As a consequence of the interactions, both components will present a different scaling behavior with the expansion than that expected from their EoS parameters. Thus, we will introduce the following effective EoS parameters that will parameterize the time evolution of the scaling of each component \cite{Tessainer1,Maroto:2025bxy} and will be useful for the rest of the work:
\begin{equation}
    \dot{\rho}_i+3\frac{\dot{a}}{a}[1+w_{\mathrm{eff},i}(a)]\,\rho_i=0.
    \label{weff_definition_gen}
\end{equation}
Now, we will present the two cases of interest and discuss the fundamental source of the effective interaction.   
\subsubsection{Shift-symmetric model}\label{sec221}
In the shift-symmetric model, the matter action is constituted by two kinetically driven fields, i.e.,
\begin{equation}
    S_\mathrm{mat}=\int \dd^4 x\sqrt{g}[H_1(Y)X_1+H_2(Y)X_2].
    \label{S_2kin_gen}
\end{equation}
 Since there is no interaction potential between the fields, the EoM for each field $\phi_i$ will simply result in
\begin{equation}
    X_i=\frac{C_{\phi_i}^2}{2H_i(Y)^2a^6}.
    \label{X_i_gen}
\end{equation}
with $C_{\phi_i}$ constant. Thus, computing the EoM for the Stueckelberg field $A^\mu$ and introducing this result yields the constraint: 
\begin{equation}
    \begin{split}
    H_1'(Y)X_1+H_2'(Y)X_2=C_g,  
    \end{split}
\end{equation}
which implies that 
\begin{equation}
    H_1'(Y)\frac{C_{\phi_1}^2}{2H_1(Y)^2}+H_2'(Y)\frac{C_{\phi_2}^2}{2H_2(Y)^2}=C_ga^6,
    \label{constraint_2kin_gen}
\end{equation}
where $C_g$ is a constant. This expression allows us to obtain $Y$ in terms of the scale factor. This expression  has contributions from both fields so that the $Y(a)$ relation will  be determined by the dynamics of the two fields. Consequently, we can view the Diff symmetry breaking as a source of a natural interaction taking place between the fields, even if they are free and there is no interaction term at the level of the action. As was shown in \cite{Tessainer1}, when one field is dominant over the other, the $Y(a)$ function will be approximately equal to that obtained in the single-field case for the dominant component, but the behavior of the subdominant field will be altered by the interaction and a wide range of phenomenological scenarios will be possible. In general, both fields will present a dynamical scaling behavior (parameterized by \eqref{weff_definition_gen}) as a consequence of the interaction, with their asymptotic single-case behavior being reached when they become dominant. For completion, the energy densities of each field will thus simply read, 
\begin{equation}
    \rho_i=X_i[H_i(Y)+YH_i'(Y)],
    \label{rho_i_shift_symmetric}
\end{equation}
with the effects of the interaction being reflected in the particular shape of the $Y(a)$ function. 

In the case of interest in which the coupling functions are power-laws of $Y$, i.e., $H_i(Y)=\lambda_iY^{\alpha_i}$, with $\lambda_i=\text{const.}$, the energy densities will simply read
\begin{equation}
    \rho_i=\lambda_i(1+\alpha_i)X_iY^{\alpha_i}.
    \label{rho_power_2kin}
\end{equation}
Thus, the asymptotic EoS parameters that depict the behavior of each component when they dominate will be 
\begin{equation}
    w_i=\frac{1-\alpha_i}{1+\alpha_i}.
    \label{w_asympt_2kin}
\end{equation}
Translating the results of \cite{Tessainer1} into the covariantized formalism allows us to obtain the following EoS functions that govern the dynamical scaling of each field, which is a result of the effective interactions: 
\begin{equation}
    w_{\mathrm{eff},i}(a)=\frac{1}{3}a\left[\frac{6}{a}+\alpha_i\frac{Y'(a)}{Y(a)}\right]-1.
    \label{weff_2kin}
\end{equation}
In particular, if one of the fields ($\phi_1$, for example) is dominant over the other, neglecting the subdominant contribution in \eqref{constraint_2kin_gen} we can see that the subdominant field will scale as \cite{Tessainer1} 
\begin{equation}
    w_{\mathrm{eff},2}\simeq1-\frac{2\alpha_2}{1+\alpha_1},
    \label{weff2kin_dom}
\end{equation}
which allows for $\phi_2$ to present a wide variety of possible behaviors, including phantom and quintessence dark energy. As shown in \cite{Tessainer1}, for fields with $w_1$, $w_2<1$ (below the asymptotic stiff fluid behavior), the energy exchange always favors the component with lower $w_i$ and the single-field domination regimes are reached.

\subsubsection{Mixed-regime model}\label{sec222}
The mixed-regime model is the second case of interest and it involves a matter action consisting of a kinetically driven field and another field dominated by its potential contribution  \cite{Maroto:2025bxy}:
\begin{equation}
    S_\mathrm{mat}=\int\dd^4 x\sqrt{g}[-H_1(Y)V_1(\phi_1)+H_2(Y)X_2].
    \label{S_mixto_gen}
\end{equation}
The EoM for the potential field will simply yield $V_1(\phi_1)=\text{const.}$, while the EoM for the kinetic component will simply result in \eqref{X_i_gen}. The EoM for the Stueckelberg field $A^\mu$ read
\begin{equation}
    \begin{split}
    H_2'(Y)X_2-H_1'(Y)V_1(\phi_1)=C_g,
    \end{split}
\end{equation}
which implies that 
\begin{equation}
    H_2'(Y)\frac{C_{\phi_2}^2}{2H_2(Y)^2a^6}-H_1'(Y)V_1(\phi_1)=C_g,
    \label{constrain_mixto}
\end{equation}
with $C_g$ a constant. This result also has contributions from both fields, resulting in an effective interaction between the fields that will depend on the specific shape of the $Y(a)$ function. As shown in \cite{Maroto:2025bxy}, both fields will present a dynamical scaling due to this interaction and will only reach their asymptotic single-case behavior when dominant. The energy densities of each field will be given by 
\begin{align}
    \rho_1&=V_1(\phi_1)[H_1(Y)-YH_1'(Y)],
    \label{rho1_mixto}\\
    \rho_2&=X_2[H_2(Y)+YH_2'(Y)].
    \label{rho2_mixto}
\end{align}
In the specific case in which the coupling functions are power-laws of $Y$, i.e., $H_i(Y)=\lambda_iY^{\alpha_i}$, with $\lambda_i=\text{const.}$, the energy densities of each field will read 
\begin{align}
    \rho_1&=\lambda_1 V_1(\phi_1)(1-\alpha_1)Y^{\alpha_1},
    \label{rho1_mixto_exp}\\
    \rho_2&=\lambda_2(1+\alpha_2)X_2Y^{\alpha_2},
    \label{rho2_mixto_exp}
\end{align}
and the asymptotic EoS parameters for $\phi_1$ and $\phi_2$ will be given by $w_1=-1$ and \eqref{w_asympt_2kin}, respectively. As obtained in \cite{Maroto:2025bxy}, the effective EoS functions that represent the dynamical scaling of each component read
\begin{align}
    w_\text{eff,1}(a)&=-\frac{1}{3}\alpha_1a\frac{Y'(a)}{Y(a)}-1,
    \label{weff1mixto}\\
    w_\text{eff,2}(a)&=\frac{1}{3}a\left[\frac{6}{a}+\alpha_2\frac{Y'(a)}{Y(a)}\right]-1.
    \label{weff2mixto}
\end{align}
In particular, the kinetic field will present stiff behavior under $\phi_1$ domination ($w_\text{eff,2}\simeq1$) and the potential field will be able to present a wide variety of possible behaviors under $\phi_2$ domination, differing considerably from a cosmological constant:
\begin{equation}
    w_\text{eff,1}\simeq\alpha_1\left(1+\frac{1-\alpha_2}{1+\alpha_2}\right)-1.
    \label{w_eff_asymp_1_kindom_powerlaw}
\end{equation}
If we leave $\alpha_1$ and $\alpha_2$ as free parameters, proceeding in an analogous way to \cite{Maroto:2025bxy}, we can infer that if $\alpha_1<0$, the potential field will behave like a phantom component under kinetic domination and the energy exchange will be in favor of it, allowing the asymptotic potential single-field potential domination regime to be reached. On the other hand, if $0<\alpha_1<1$, the potential field will lose energy and both fields will end up tracking each other in the future. We can illustrate this tracking in an analogous way to \cite{Maroto:2025bxy} by writing \eqref{constrain_mixto} asymptotically assuming  $Y(a)\sim ka^{-\beta}$ with $\beta>0$\footnote{As shown in \cite{Maroto:2025bxy}, $Y(a)$ is a decreasing function.} for $a\rightarrow\infty$:
\begin{equation}
    \frac{1+\alpha_1 C_1}{\alpha_2k^{\alpha_2+1}}\frac{a^{\beta\alpha_2+\beta}}{a^6}-\alpha_1C_1k^{\alpha_1-1}\frac{a^\beta}{a^{\beta\alpha_1}}=1,
    \label{constr_asymptotic_mixto}
\end{equation}
where $C_1\equiv\lambda_1 V_1(\phi_1)$ and where $C_2=(1+\alpha_1C_1)/\alpha_2$ was fixed by using the normalization $Y_0=Y(a_0=1)=1$. Since $\phi_1$ does not reach its asymptotic single-field domination regime in this case, \eqref{constr_asymptotic_mixto} will be satisfied for $a\rightarrow\infty$ just if 
\begin{equation}
\beta=6/(\alpha_1+\alpha_2),
\label{beta_tracking}
\end{equation}
which results in the expected tracking, i.e.,
\begin{equation}
    w_{\mathrm{eff},1}=w_{\mathrm{eff},2}=w_\mathrm{track}\equiv\frac{\alpha_1-\alpha_2}{\alpha_1+\alpha_2}.
    \label{w_tracking}
\end{equation}

\section{Cosmological perturbations: the single-field case}\label{sec3}
In this section, we develop the formalism of scalar perturbations in single-field TDiff theories on a flat FLRW background. We analyze several cases of interest, including the kinetic and potential domination regimes, as well as the most general case.  We will work in the covariantized approach so that we can use the standard cosmological perturbation theory.

Taking into account that the gravitational sector of the action is left unchanged (see \eqref{S_tot}), the most general scalar-perturbed flat FLRW spacetime reads
\begin{equation}
    \dd s^2=a(\eta)^2[(1+2\Phi)\dd\eta^2-2B_{,i}\dd x^i\dd\eta-((1-2\Psi)\delta_{ij}+2E_{,ij})\dd x^i\dd x^j],
    \label{FLRW_pert_escalar}
\end{equation}
where $\Phi$, $\Psi$, $B$ and $E$ are the respective scalar perturbations. If we perform a general gauge transformation $x^\mu\rightarrow\hat{x}^\mu=x^\mu+\xi^\mu$, assuming that it only affects the perturbation and that $\xi^\mu=(\xi^0,\xi^i)$ is a first order contribution, we have that the metric and EMT transform as
\begin{align}
    \delta\hat{g}_{\mu\nu}&=\delta g_{\mu\nu}-\mathfrak{L}_\xi(g_{\mu\nu}^{(0)}),
    \label{delta_g_diff}\\
    \delta\hat{T}_{\mu\nu}&=\delta T_{\mu\nu}-\mathfrak{L}_\xi(T_{\mu\nu}^{(0)}),
    \label{delta_EMT_diff}
\end{align}
where the superscript $(0)$ will refer to background quantities from now on  and $\mathfrak{L}_\xi$ denotes the Lie derivative along $\xi^\mu$. The scalar perturbations will obey the following gauge transformation rules:
\begin{align}
    \hat{\Phi}&=\Phi-\mathcal{H}\xi^0-\dot{\xi}^{0},
    \label{phi_gauge}\\
    \hat{\Psi}&=\Psi+\mathcal{H}\xi^0,
    \label{psi_gauge}\\
    \hat{B}&=B+\xi^0-\dot{\xi},
    \label{B_gauge}\\
    \hat{E}&=E-\xi,
    \label{E_gauge}
\end{align}
where we decomposed the spatial part of the transformation as the sum of a scalar gradient and a pure vector contribution, i.e., $\xi^i=\delta^{ij}\xi_{,j}+\Bar{\xi}^i$, with $\mathcal{H}=\dot{a}/a$ the conformal Hubble rate. Accordingly, the gauge-invariant Bardeen potentials will read
\begin{align}
    \Phi_B&=\Phi+\mathcal{H}(B-\dot{E})+\frac{\partial}{\partial\eta}(B-\dot{E}),\\
   \Psi_B&=\Psi-\mathcal{H}(B-\dot{E}).
\end{align}
With regards to the matter sector, the perturbations of a perfect fluid with an EMT given by \eqref{EMT_perfect_fluid} read
\begin{align}
    \delta T^0_{\;0}&=\delta\rho,
    \label{delta_T00_gen}\\
   \delta T^i_{\;j}&=-\delta p\,\delta^i_j,
    \label{delta_Tij_gen}\\
    \delta T^0_{\;i}&=-(\rho^{(0)}+p^{(0)})(B+v)_{,i}\equiv-\delta q_{,i},
    \label{delta_T0i_gen}
\end{align}
where $\delta q$ will be referred to as the momentum perturbation and $v^i$ is the perturbed three-velocity, which can also be expressed as the sum of a scalar gradient and a pure vector contribution: $v^i=\delta^{ij}v_{,j}+\Bar{v}^i$. These matter perturbations have the following gauge transformation rules:
\begin{align}
    \delta\hat{\rho}=&\delta\rho-\dot{\rho}^{(0)}\xi^0,
    \label{delta_rho_gauge}\\
   \delta\hat{p}=&\delta p-\dot{p}^{(0)}\xi^0,
    \label{delta_p_gauge}\\
    \delta\hat{q}=&\delta q+(\rho^{(0)}+p^{(0)})\xi^0,
\end{align}
and their specific shape depends on the particular theory underlying the description of the matter content. Thus, the perturbed Einstein equations up to the first order $\delta G^{\mu\nu}=8\pi G\delta T^{\mu\nu}$ yield:
\begin{align}
    &3\mathcal{H}^2\Phi+3\mathcal{H}\dot{\Psi}-\nabla^2[\Psi-\mathcal{H}(B-\dot{E})]=-4\pi Ga^2\delta\rho,
    \label{EEP1}\\
    &\dot{\Psi}+\mathcal{H}\Phi=-4\pi Ga^2\delta q,
    \label{EEP2}\\
    &\ddot{\Psi}+2\mathcal{H}\dot{\Psi}+\mathcal{H}\dot
    \Phi+\mathcal{H}^2\Phi+2\mathcal{\dot{H}}\Phi=4\pi Ga^2\delta p,
    \label{EEP3}\\
    &\Phi-\Psi+\mathcal{H}(B-\dot{E})+\frac{1}{a}\frac{\partial}{\partial\eta}(aB-a\dot{E})=0,
    \label{EEP4}
\end{align}
and the perturbed EMT conservation equations read 
\begin{align}
   \delta\dot{\rho}&+3\mathcal{H}(\delta\rho+\delta p)-3(\rho^{(0)}+p^{(0)})\dot{\Psi} \nonumber  \\ 
    &=-\nabla^2[\delta q-(\rho^{(0)}+p^{(0)})(B-\dot{E})],
    \label{EMT_cons_pert}
    \end{align}
    \begin{align}
    \delta\dot{q}+4\mathcal{H}\delta q+\nabla^2\delta p+(\rho^{(0)}+p^{(0)})\nabla^2\Phi=0.
    \label{EMT_cons_pert_2}
\end{align}
Having presented this formalism, we will now first discuss the EMT perturbations of single-field TDiff theories in the most general scenario, as well as in the kinetic and potential domination regimes.
\subsection{General case}
We will start with the discussion of the general single-field case with both a kinetic and a potential term. In the general case, the action and the EMT at the background level will simply be given by \eqref{S_1scalar_gen_cov} and \eqref{EMT_perfect_fluid}. Since the field is homogeneous at the background level, its perturbations will read
\begin{equation}
    \phi(x^\mu)=\phi^{(0)}(\eta)+\delta\phi(x^\mu),
    \label{phi_pert}
\end{equation}
and the first-order perturbations of the EMT will read:
\begin{align}
    \delta\rho&=\delta X(H_K+YH_K)+\delta Y(XYH_K''+2XH_K'-YH_V''V)\nonumber  \\ 
    &+\delta\phi V'(H_V-YH_V'),
    \label{delta_rho_1gen1}\\
    \delta p&=\delta X(H_K-YH_K)+\delta Y(YH_V''V-XYH_K'')  \nonumber \\ 
    &-\delta\phi V'(H_V-YH_V'),
    \label{delta_p_1gen1}\\
    \delta q&=-H_K\frac{1}{a^2}\phi^{(0)}{'}\delta\phi,
    \label{delta_q_1gen1}
\end{align}
where 
\begin{equation}
    \delta X=\frac{1}{a^2}\dot{\phi}{^{(0)}}\delta\dot{\phi}-\frac{\Phi}{a^2}[\dot{\phi}{^{(0)}}]^2;
    \label{deltaX}
\end{equation}
and where from now on we will denote $H_i\equiv H_i(Y)$ and $V\equiv V(\phi)$ to avoid fuzzy notation. It is worth noting, however, that $\delta Y$ is not an independent perturbation, since it can be written in terms of the other perturbations when taking into account the EoM for the Stueckelberg field, which in this case yields
\begin{equation}
    H_K'(Y)X-H_V'(Y)V(\phi)=\text{const.}
    \label{EoM_Amu_gen}
\end{equation}
Thus, $\delta Y$ can be expressed as 
\begin{equation}
    \delta Y=\frac{H_V'(Y)V'(\phi)\delta\phi-H_K'(Y)\delta X}{H_K''(Y)X-H_V''(Y)V(\phi)},
    \label{delta_Y_gen}
\end{equation}
and we can write the EMT perturbations in terms of $\delta X$ and $\delta\phi$:
\begin{align}
    \delta\rho&=\delta X\frac{H_K H_V''V+2H_K'^2X-H_KH_K''X}{H_V''V-H_K''X}\nonumber \\ 
    &+\delta\phi V'\frac{H_K''H_VX+2H_K'H_V'X-H_VH_V''V}{H_K''X-H_V''V},&
    \label{delta_rho_1gen}\\
    \delta p&=H_K\delta X-H_V V'\delta\phi,&
    \label{delta_p_1gen}\\
    \delta q&=-H_K\frac{1}{a^2}\phi^{(0)}{'}\delta\phi,&
    \label{delta_q_1gen}
\end{align}
Having computed all these perturbations, we can now obtain the speed of sound. In fact, we have three matter perturbations ($\delta\rho$, $\delta p$ and $\delta q$), but there are only two field perturbations ($\delta X$ and $\delta\phi$). Therefore, only two of said perturbations will be independent and we could write \cite{Unnikrishnan:2024agf}
\begin{equation}
    \delta p=c_s^2\,\delta\rho+3\mathcal{H}(c_a^2-c_s^2)\,\delta q,
    \label{delta_p_1field_gen}
\end{equation}
where the coefficients of each term were chosen so that the expression has the same form in every gauge and where the $c_s^2$ coefficient is the speed of sound at which perturbations propagate in the sub-Hubble limit \cite{Unnikrishnan:2024agf}. Notice that we can write 
\begin{equation}
    c_s^2\equiv\frac{\delta p_c}{\delta\rho_c}
    \label{cs^2_def_general}
\end{equation}
where $\delta p_c$ and $\delta \rho_c$ are  the pressure and energy density perturbations evaluated in the comoving gauge 
 ($\delta q=0$, i.e., $\delta\phi=0$). On the other hand,  $c_a^2=p^{(0)}{'}/\rho^{(0)'}$ is the adiabatic speed of sound. We can thus also express the pressure perturbation in the following way: 
\begin{equation}
    \delta p=c_a^2\,\delta\rho+\delta p_\text{nad},
    \label{delta_p_1field_nad}
\end{equation}
where 
\begin{equation}
\delta p_\text{nad}=(c_s^2-c_a^2)(\delta\rho-3\mathcal{H}\delta q)
\label{delta_pnad_1field}
\end{equation}
denotes the gauge-invariant non-adiabatic pressure perturbation, which vanishes if $c_s^2=c_a^2$ and in such cases the perturbations will be adiabatic and propagate as a wave with speed $c_s^2=c_a^2$ in the sub-Hubble limit. The speed of sound can also be defined in a gauge-invariant way introducing the following gauge invariants \cite{Unnikrishnan:2024agf}:
\begin{equation}
    c_s^2=\frac{\delta p_c}{\delta \rho_c}\equiv\frac{\delta p-3\mathcal{H}c_a^2\delta q}{\delta\rho-3\mathcal{H}\delta q},
    \label{cs_gauge_inv}
\end{equation}
 Thus, going to the comoving gauge in \eqref{delta_rho_1gen} and 
 \eqref{delta_p_1gen}, we obtain for the speed of sound:
\begin{equation}
    c_s^2=\frac{H_KH_V''V-H_KH_K''X}{H_KH_V''V+2H_K'^2X-H_KH_K''X},
    \label{cs2_gen_1field}
\end{equation}
which is equivalent to the expression obtained in \cite{Dario_cov}. Thus we find that in general the perturbations will not be  adiabatic and thus $c_s^2\neq c_a^2$ ($\delta p_\text{nad}\neq 0$), as shown in \cite{Dario_cov}, which could be expected since the energy density and pressure depend on two variables, namely $Y$ and $\phi$.
\subsection{Potential domination regime}
We will now analyze the potential domination regime, in which $X\rightarrow0$. We can study this case by taking this limit in the above computed general expression \eqref{cs2_gen_1field}. This yields
\begin{equation}
    c_s^2=1,
    \label{cs2_1field_pot}
\end{equation}
and thus perturbations propagate at the speed of light, which can be justified since $\phi$ is not an exact cosmological constant and has energy and pressure perturbations. Notice that the adiabatic speed of sound should be computed to first order in the slow-roll approximation in this regime (see \cite{Maroto:2025wtg}).

\subsection{Kinetic domination regime}
Lastly, we will analyze the case in which the field is dominated by its kinetic term. The action in this case reads 
\begin{equation}
    S_\mathrm{mat}=\int \dd^4 x\sqrt{g}H_K(Y)X,
    \label{S_1kin}
\end{equation}
Thus, performing the kinetic limit in \eqref{cs2_gen_1field} will have that 
\begin{equation}
    c_s^2=\frac{H_K(Y)H_K''(Y)}{H_K(Y)H_K''(Y)-2H_K'(Y)^2},
    \label{c_s_1kin}
\end{equation}
which is equivalent to the result obtained in \cite{Dario_cov} and it is also equal to the adiabatic speed of sound calculated in the same reference (this could be expected since the energy density and pressure only depend on one variable). Thus, the pressure perturbations in single-field theories with a kinetically driven field will be adiabatic and $c_s^2=c_a^2$. 
\section{Cosmological perturbations: the multi-field case}\label{sec4}
In this section, we will develop the formalism of scalar perturbations in multi-field TDiff theories on a flat cosmological background and we will discuss the different terms that appear in the pressure perturbations. We will focus on the study of the previously discussed shift-symmetric and mixed-regime models, and we will compute the respective speed of sound, discussing the adiabaticity in each case. We will also regard the stability of each model and discuss the power-law coupling case in further detail for its phenomenological interest. 

Before proceeding with the calculations for the specific models of study, we will perform a theoretical analysis of our degrees of freedom when it comes to the perturbations and discuss the different contributions that will appear in the pressure perturbation in general. On the one hand, at the field level,  perturbations will only depend on four independent terms, namely, $\delta X_1$, $\delta X_2$, $\delta\phi_1$ and $\delta\phi_2$ (see \eqref{delta_rho_1gen1}, \eqref{delta_p_1gen1} and \eqref{delta_q_1gen1}). On the other hand, at the EMT level,  we have $\delta p$, as well as $\delta\rho$ and $\delta q$. However, we can also use $\delta\rho_i$ and $\delta q_i$ with $i=1,2$, in order to introduce relative energy and momentum perturbations through the following gauge-invariant quantities \cite{Unnikrishnan:2024agf,Wands:2007bd}:
\begin{eqnarray}
    \delta\rho_r&\equiv&\rho^{(0)}\left(\frac{\delta\rho_1}{\rho_1^{(0)}{'}(a)}-\frac{\delta\rho_2}{\rho_2^{(0)}{'}(a)}\right),
    \label{delta rhominus}\\
    \delta q_r&\equiv&\mathcal{H}\left(\frac{\delta q_1}{\rho_1^{(0)}+p_1^{(0)}}-\frac{\delta q_2}{\rho_2^{(0)}+p_2^{(0)}}\right),
    \label{delta qminus}
\end{eqnarray}
where the coefficients multiplying the parentheses have been chosen that way so that the respective perturbation coefficients are dimensionless. Thus, inspired by the single-field analysis and taking into account that the amount of independent perturbations is four, we can write, in a gauge invariant way:
\begin{equation}
    \delta p=c_s^2\,\delta\rho+3\mathcal{H}(c_a^2-c_s^2)\,\delta q+\delta\Tilde{p},
    \label{deltap_2kin_step1}
\end{equation}
where 
\begin{equation}
    \delta\Tilde{p}=\kappa\,\delta \rho_r+\sigma\, \delta q_r
    \label{delta p tilde_2kin}
\end{equation}
corresponds to the contribution associated to the presence of multiple components. In these expressions, $c_s^2$ denotes the effective speed of sound, while $\kappa$ and $\sigma$ are non-adiabatic coefficients. Consequently, the general expression for the pressure perturbation in a multi-field model will read 
\begin{align}
    \delta p&=c_s^2\,\delta\rho+3\mathcal{H}(c_a^2-c_s^2)\,\delta q +\kappa\,\rho^{(0)}\left(\frac{\delta\rho_1}{\rho_1^{(0)}{'}}-\frac{\delta\rho_2}{\rho_2^{(0)}{'}}\right) \nonumber\\ 
    &+\sigma\,{\mathcal{H}}\left(\frac{\delta q_1}{\rho_1^{(0)}+p_1^{(0)}}-\frac{\delta q_2}{\rho_2^{(0)}+p_2^{(0)}}\right),
    \label{delta p gen 2kin}
\end{align}
where from now on we will denote $\rho_i^{(0)}{'}\equiv\rho_i^{(0)}{'}(a)$ for simplicity. As we can see, if $c_s^2$ is equal to $c_a^2$, this will no longer guarantee that the perturbations are adiabatic, as there could be non-zero relative perturbations between the fields. However, it can be shown that perturbations propagate as waves with speed $c_s^2$ in the sub-Hubble regime \cite{Unnikrishnan:2024agf}.

\subsection{Shift-symmetric model}\label{sec4p1}
As introduced previously, the shift-symmetric model involves two scalar fields driven by their kinetic term, i.e., 
\begin{equation}
    S_\text{mat}=\int \dd^4 x\sqrt{g}[H_1(Y)X_1+H_2(Y)X_2],
    \label{S_2kin_denuevo}
\end{equation}
 where the fields equations of motion imply $X_i={C_i}/(H_i(Y)^2a^6),$ with $C_i\equiv C_{\phi_i}^2/2$. As the EoM for the Stueckelberg field $A^\mu
 $ indicate (see \eqref{constraint_2kin_gen}), $H_1'X_1+H_2'X_2$ will be a constant and solving this constraint allows us to obtain $Y(a)$, which will affect the energy densities and trigger an effective interaction between the fields, even if there is no interacting term in the action. Perturbing this constraint \eqref{constraint_2kin_gen} allows us to obtain $\delta Y$ in terms of $\delta X_1$ and $\delta X_2$:
\begin{equation}
    \delta Y=-\frac{H_1'\delta X_1+H_2'\delta X_2}{X_1H_1''+X_2H_2''},
    \label{delta Y_2kin}
\end{equation}
which has contributions from both fields and will be indicative of the interactions at the perturbative level. 

We will now apply the above discussed formalism to this shift-symmetric case and compute the effective speed of sound $c_s^2$. Since there is an effective interaction taking place between the fields given by the specific shape of $Y$, we will not be able to express $\delta p_1$ and $\delta p_2$ as in \eqref{delta_p_1field_gen} and we must compute $c_s^2$ and the  rest of the parameters directly from \eqref{delta p gen 2kin}. As discussed previously, using the constraint and \eqref{delta Y_2kin} we can express the energy and pressure perturbations in terms of $\delta X_1$ and $\delta X_2$ in the following way
\begin{align}
    \delta p_1=&\mathcal{A}\,\delta X_1+\mathcal{B}\,\delta X_2,
    \label{deltap1compact}\\
    \delta p_2=&\mathcal{C}\,\delta X_1+\mathcal{D}\,\delta X_2,
    \label{deltap2compact}\\
    \delta\rho_1=&\mathcal{E}\,\delta X_1+\mathcal{F}\,\delta X_2,
    \label{deltarho1compact}\\
    \delta\rho_2=&\mathcal{G}\,\delta X_1+\mathcal{J}\,\delta X_2,
\end{align}
where the coefficients are written in detail in section \ref{Appendix} (appendix). Thus, by equating the coefficients that multiply $\delta X_1$ and $\delta X_2$ in \eqref{delta p gen 2kin} we can obtain the speed of sound and the $\kappa$ coefficient:
\begin{align}
    c_s^2&=\frac {1}{(\mathcal{FG}-\mathcal{EJ})(\rho_1^{(0)}{'}+\rho_2^{(0)}{'})}\left[\mathcal{B}(\mathcal{G}\rho_1^{(0)}{'}-\mathcal{E}\rho_2^{(0)}{'}) \right. \nonumber\\ 
    &\left. +\mathcal{D}(\mathcal{G}\rho_1^{(0)}{'}-\mathcal{E}\rho_2^{(0)}{'})-(\mathcal{A}+\mathcal{C})(\mathcal{J}\rho_1^{(0)}{'}-\mathcal{F}\rho_2^{(0)}{'})\right],
    \label{cs_2kin_gen}\\
    \kappa&=\frac{\rho_1^{(0)}{'}\rho_2^{(0)}{'}}{\rho^{(0)}}\frac{\mathcal{A+\mathcal{C}}-c_s^2(\mathcal{E}+\mathcal{G})}{\mathcal{E}\rho_2^{(0)}{'}-\mathcal{G}\rho_1^{(0)}{'}},
    \label{kapparho_2kingen}
\end{align}
where $\kappa$ will generally be non-zero and there will thus be a non-adiabatic contribution to the pressure perturbation involving relative intrinsic energy density perturbations. With regards to the adiabatic speed of sound, it can be checked that, for the shift-symmetric case,
\begin{equation}
c_a^2=\frac{p^{(0)}{'}}{\rho^{(0)}{'}} =c_s^2,
\label{check_cs_ca_2kin}
\end{equation}
as it was checked using the software \url{Mathematica}\footnote{\url{https://www.wolfram.com/mathematica/}} after expanding the $\rho_i^{(0)}{'}$ in \eqref{cs_2kin_gen}. Thus, the term proportional to $\delta q$ in \eqref{delta p gen 2kin} will vanish. On the other hand, 
since $\delta q_i\propto \delta\phi_i$ and the l.h.s. of \eqref{delta p gen 2kin} does not involve $\delta\phi_i$ terms in this case, the contribution proportional to the relative momentum perturbation $\delta q_r$ must vanish. However, since $\delta q_r\neq 0$ in general and the coefficients in \eqref{delta p gen 2kin} are gauge invariant quantities, we conclude that 
\begin{equation}
\sigma=0,
\label{sigma=0_2kin}
\end{equation}
 Therefore, we will have a model with $c_s^2=c_a^2$ but with a non-adiabatic pressure perturbation contribution related to relative energy density perturbations. 
\subsubsection{Power-law coupling functions}
We will now analyze in detail the case in which both coupling functions are power-laws of $Y$, i.e., $H_i(Y)=\lambda_iY^{\alpha_i}$. In this case, substituting these functions in \eqref{cs_2kin_gen} yields
\begin{align}
    c_s^2
    =\frac{c_1^2\,\rho_1^{(0)}{'}+c_2^2\,\rho_2^{(0)}{'}}{\rho_1^{(0)}{'}+\rho_2^{(0)}{'}},
    \label{cs_2kin_power_gen}
\end{align}
where $c_i^2$ are the individual speeds of sound of each field in the power-law case, where it can be checked that $c_i^2=w_i$ is satisfied, with $w_i$ given by \eqref{w_asympt_2kin}. We can rewrite this expression using the conservation equations \eqref{cons1Q} and \eqref{cons2Q}, which leads to
\begin{equation}
    c_s^2=\frac{(\rho_1^{(0)}+p_1^{(0)})\,c_1^2+(\rho_2^{(0)}+p_2^{(0)})\,c_2^2}{\rho^{(0)}+p^{(0)}}-\frac{Q\,(c_2^2-c_1^2)}{(\rho_1^{(0)}{'}+\rho_2^{(0)}{'})}.
    \label{cs_2kin_powerlaw_Q}
\end{equation}
This expression allows us to separately distinguish the two contributions for the speed of sound. On the one hand, the first term is the \textit{free} contribution, which only depends on the sum of the individual quantities of each field. On the other hand, the second term is associated with the interaction, as it is proportional to the interaction kernel $Q$, and it vanishes if $c_1^2=c_2^2$, i.e., if both fields have the same couplings. With regard to $\kappa$, we obtain 
\begin{equation}
    \kappa=\frac{-2(\alpha_1-\alpha_2)}{(1+\alpha_1)(1+\alpha_2)\,\rho^{(0)}}\frac{\rho_1^{(0)}{'}\rho_2^{(0)}{'}}{(\rho_1^{(0)}{'}+\rho_2^{(0)}{'})},
    \label{kappa_2kin_power_law}
\end{equation}
which when taking into account the conservation equations reads
\begin{align}
    \kappa&=\frac{-2(\alpha_1-\alpha_2)\rho{^{(0)}}^{-1}}{(1+\alpha_1)(1+\alpha_2)(\rho_1^{(0)}{'}+\rho_2^{(0)}{'})}\left[\frac{9}{a^2}(\rho_1^{(0)}+p_1^{(0)})(\rho_2^{(0)}+p_2^{(0)})\right. \nonumber \\ 
    &\left. +\frac{3Q}{a^2\mathcal{H}}(\rho_1^{(0)}+p_1^{(0)}-\rho_2^{(0)}-p_2^{(0)})-\frac{Q^2}{a^2\mathcal{H}^2}\right],
    \label{kappa_2kin_power_law_Q}
\end{align}
and thus $\kappa\neq0$ unless $\alpha_1=\alpha_2$, i.e., when both fields have the same couplings. As we can appreciate from these results, since the single-field domination regimes are reached in this model \cite{Tessainer1}, $c_s^2$ will tend to the respective $c_i^2=w_i$ in said limits, with the effects of the interactions being mostly reflected in the intermediate regime, in which the terms proportional to $Q$ become more relevant. On the other hand, as we can see from \eqref{kappa_2kin_power_law_Q}, $\kappa$ can be expected to tend to zero in the single-field domination limits and to grow in the intermediate regimes as $Q$ becomes stronger.

\begin{figure}[h]
     \centering
     \includegraphics[scale=0.55]{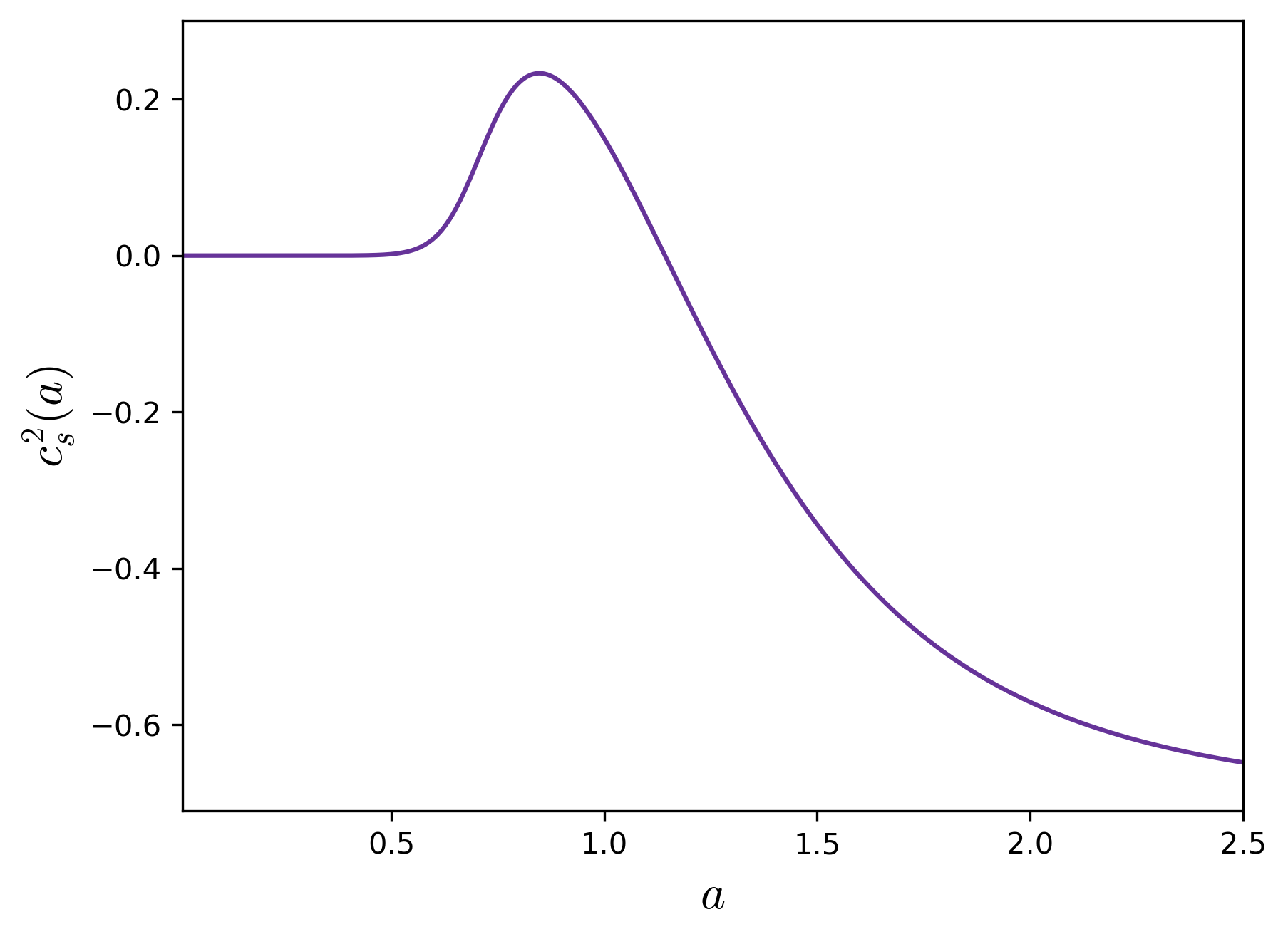}
     \caption{$c_s^2$ in terms of $a$ for $\alpha_1=1$ and $\alpha_2=5.74$.}
     \label{cs22kin}
\end{figure}

\begin{figure}[h]
    \centering
     \includegraphics[scale=0.55]{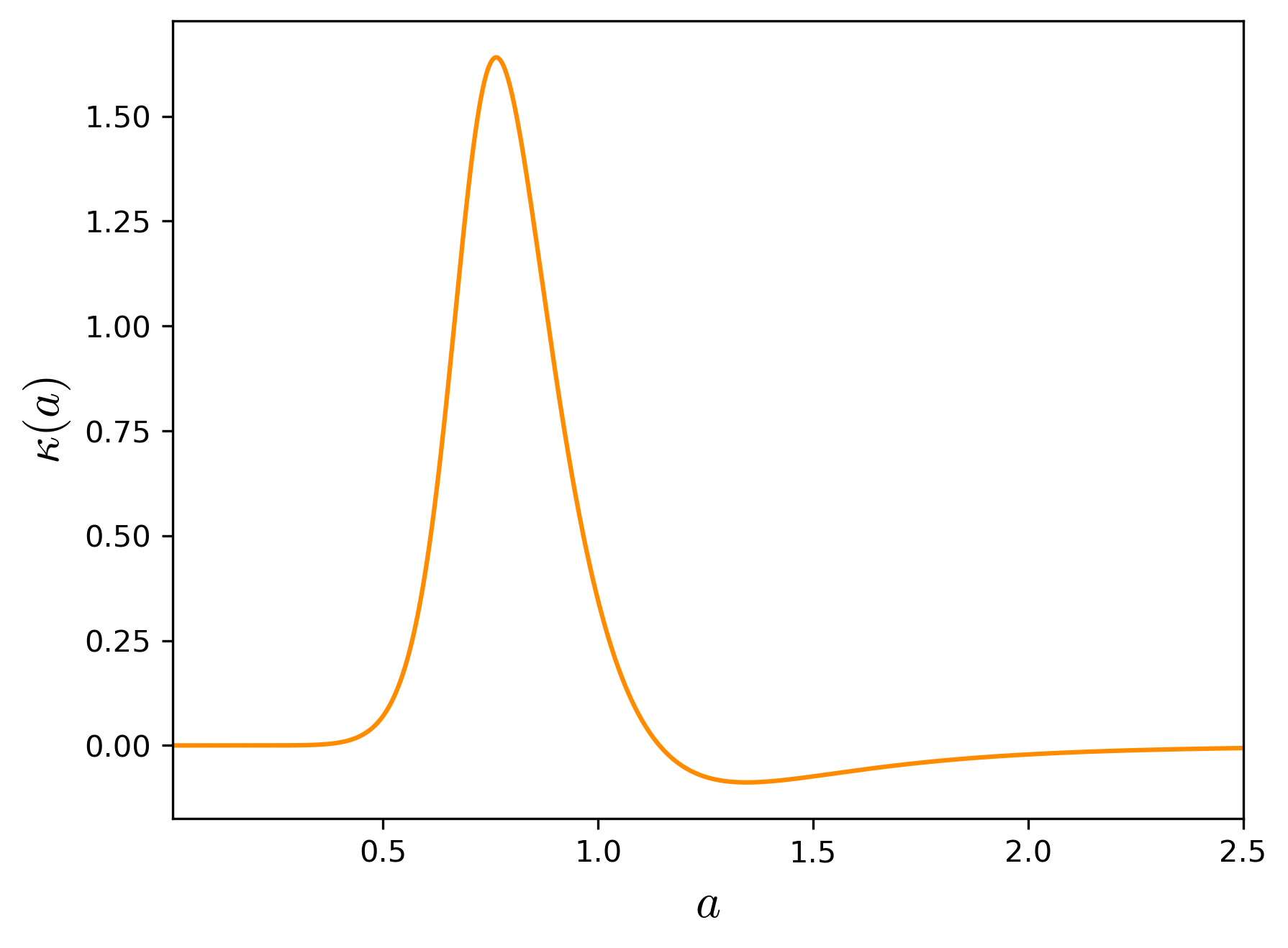}
     \caption{$\kappa$ in terms of $a$ for $\alpha_1=1$ and $\alpha_2=5.74$.}
     \label{kappa2kin}
\end{figure}

We will now discuss the stability of this models and their phenomenology. With this in mind, we will obtain the evolution of $c_s^2$ and $\kappa$. Firstly, we shall express both quantities in terms of physical parameters. Thus, we will introduce the density parameters of each field as
\begin{equation}
    \Omega_i=\frac{\rho_{i,0}^{(0)}}{\rho_\mathrm{crit}}=\frac{C_i}{\lambda_i\,\rho_\mathrm{crit}}(1+\alpha_i),
    \label{Omega_2kin}
\end{equation}
where $\rho_\mathrm{crit}$ is the critical density and $C_i\equiv C_{\phi_i^2}/2$. Taking the derivative of the constraint \eqref{constraint_2kin_gen} with respect to $a$ allows us to express $Y'(a)$ in terms of $Y(a)$ and the parameters of the model. Thus, we obtain that 
\begin{align}
        c_s^2&=\frac{1}{(\alpha_1\Omega_1Y^{\alpha_1}+\alpha_2\Omega_2Y^{\alpha_2})\left(\frac{\Omega_1Y^{\alpha_2}}{1+\alpha_1}+\frac{\Omega_2Y^{\alpha_1}}{1+\alpha_2}\right)} \nonumber \\ & \times\left[-\frac{Y^{2\alpha_2}(-1+\alpha_1)\alpha_1\Omega_1^2}{(1+\alpha_1)^2}-\frac{Y^{2\alpha_1}(-1+\alpha_2)\alpha_2\Omega_2^2}{(1+\alpha_2)^2}\right. \nonumber\\ & \left. +\frac{Y^{\alpha_1+\alpha_2}(\alpha_1^2+\alpha_1+\alpha_2-4\alpha_1\alpha_2+\alpha_2^2)\Omega_1\Omega_2}{(1+\alpha_1)(1+\alpha_2)}\right],
    \label{cs2_2kin_power_physical}
\end{align}
and
\begin{align}
        \kappa&=-\frac{12(\alpha_1-\alpha_2)\Omega_1\Omega_2(\Omega_1Y^{-\alpha_1}+\Omega_2Y^{-\alpha_2})^{-1}}{a(1+\alpha_1)^2(1+\alpha_2)^2(Y^{\alpha_2}\alpha_1\Omega_1+Y^{\alpha_1}\alpha_2\Omega_2)^2}   \\ 
        &\times 
        \frac{[Y^{\alpha_2}\alpha_1(1+\alpha_1-\alpha_2)(1+\alpha_2)\Omega_1+Y^{\alpha_1}(1+\alpha_1)\alpha_2\Omega_2]}{[Y^{\alpha_2}(1+\alpha_2)\Omega_1+Y^{\alpha_1}(1+\alpha_1)\Omega_2]} \nonumber \\ 
        &\times [Y^{\alpha_1}(1+\alpha_1)(-1+\alpha_1-\alpha_2)\alpha_2\Omega_2-Y^{\alpha_2}\alpha_1(1+\alpha_2)\Omega_1].\nonumber
    \label{kappa_power_law_physical}
\end{align}
If we plot both $c_s^2$ and $\kappa$  for certain values of the parameters $\alpha_1$, $\alpha_2$\footnote{We chose to use the best-fitting values of $\alpha_i$ for the dark sector described in \cite{Tessainer1}, and $C_2$ can be fixed from the normalization condition $Y_0=1$.} and $C_1$, we obtain the results shown in Fig.\ref{cs22kin} and Fig.\ref{kappa2kin} by numerically solving the constraint \eqref{constraint_2kin_gen}.
Thus, from these results we can infer that $c_s^2$ will smoothly transition from $w_1=0$ to $w_2=-0.703$, i.e. from  the larger $w_i$ to the lower one, presenting a peak in the intermediate regime when the interaction is stronger. This peak will also be higher the greater the difference $c_2^2-c_1^2$. The $\kappa$ coefficient will be approximately zero in the single-field domination regimes, but will grow during in the intermediate regime as a consequence of the interaction. On the other hand, in the asymptotic regimes $\delta p\simeq c_a^2\delta \rho$ and the evolution will approximately be adiabatic. When it comes to stability, we can see that these models will be unstable if one of the $w_i$ is lower than zero, but the instability could occur in the future by fixing the parameter $C_1$ so that $c_s^2>0$ before the current time $a=1$. When it comes to a hypothetical description of the dark sector, this could present a problem when dark energy is asymptotically dominant in the future. The peak in $c_s^2$ could also be somewhat problematic, as it results in $c_s^2$ deviating from zero in a noticeable way, which could affect the matter power spectrum \cite{Xu:2011bp}. However, although these models could be problematic for the description of dynamical dark energy, they could be useful for describing other scenarios such as dark radiation-dark matter interactions as we will see in the next section.

\subsection{Mixed-regime model}\label{sec4p2}
As we previously discussed, the action for the mixed regime model involves a kinetically driven field and a potentially driven field, i.e.,
\begin{equation}
    S_\mathrm{mat}=\int \dd^4x\sqrt{g}(-H_1(Y)V(\phi_1)+H_2(Y)X_2),
    \label{S_mixto2}
\end{equation}
From the field equations, we find  $X_2={C_2}/{(H_2(Y)^2a^6)}$. 
On the other hand, looking at the EoM for the Stueckelberg field $A^\mu$ (see \eqref{constrain_mixto}), $H_2'X_2-H_1'V_1=\mathrm{const.}$
, and solving this equation will allow us to obtain $Y(a)$. Perturbing this condition and defining $V'\equiv \dd V/\dd\phi_1$ allows us to obtain $\delta Y$ in terms of $\delta\phi_1$ and $\delta X_2$:
\begin{equation}
    \delta Y=\frac{H_1'V'\delta\phi_1-H_2'\delta X_2}{H_2''X_2-H_1''V},
    \label{deltaY_mixto}
\end{equation}
 which has contributions from both fields and reflects the interaction at the perturbative level. When it comes to the pressure perturbation $\delta p$ and its contributions, it can be checked that the analysis is analogous to the shift-symmetric case. Thus, following the previous reasoning and \cite{Wands:2007bd}, the general contribution for the pressure perturbation will have the same structure as \eqref{delta p gen 2kin}, involving two terms related to relative energy and momentum perturbations. 

With this in mind, we can calculate the speed of sound $c_s^2$ directly from the pressure perturbation \eqref{delta p gen 2kin}, expressing the individual energy and pressure perturbations in terms of $\delta\phi_1$ and $\delta X_2$:
\begin{align}
    \delta p_1=&\hat{\mathcal{A}}\,\delta\phi_1+\hat{\mathcal{B}}\,\delta X_2,
    \label{deltap1compact_mixto}\\
    \delta p_2=&\hat{\mathcal{C}}\,\delta\phi_1+\hat{\mathcal{D}}\,\delta X_2,
    \label{deltap2compact_mixto}\\
    \delta\rho_1=&\hat{\mathcal{E}}\,\delta \phi_1+\hat{\mathcal{F}}\,\delta X_2,
    \label{deltarho1compact_mixto}\\
    \delta\rho_2=&\hat{\mathcal{G}}\,\delta\phi_1+\hat{\mathcal{J}}\,\delta X_2,
\end{align}
where the coefficients can be found in section \ref{Appendix} (appendix). Hence, equating the coefficients that multiply $\delta\phi_1$ and $\delta X_2$ in \eqref{delta p gen 2kin} we can compute the speed of sound and the $\kappa$ coefficient:  
\begin{align}
    c_s^2&=\frac{(\hat{\mathcal{A}}+\hat{\mathcal{C}})\hat{\mathcal{J}}\rho_1^{(0)}{'}-\hat{\mathcal{D}}(\hat{\mathcal{G}}\rho_1^{(0)}{'}+\hat{\mathcal{A}}\rho_2^{(0)}{'})}{(\hat{\mathcal{B}}\hat{\mathcal{G}}-\hat{\mathcal{A}}\hat{\mathcal{J}})(\rho_1^{(0)}+\rho_2^{(0)})}\nonumber \\ 
    &+\frac{\hat{\mathcal{B}}(-\hat{\mathcal{G}}\rho_1^{(0)}{'}+\hat{\mathcal{C}}\rho_2^{(0)}{'})}{(\hat{\mathcal{B}}\hat{\mathcal{G}}-\hat{\mathcal{A}}\hat{\mathcal{J}})(\rho_1^{(0)}+\rho_2^{(0)})},
    \label{cs2_mixto}\\
    \kappa&=-\frac{\rho_1^{(0)}{'}\rho_2^{(0)}{'}}{\rho^{(0)}}\frac{\hat{\mathcal{A}}+\hat{\mathcal{C}}-c_s^2(\hat{\mathcal{G}}-\hat{\mathcal{A}})}{\hat{\mathcal{A}}\rho_2^{(0)}{'}+\hat{\mathcal{G}}\rho_1^{(0)}{'}}.
    \label{kappa_mixto}
\end{align}

As in the shift-symmetric case, $\kappa$ will be generally non-zero and there will also be a non-adiabatic contribution of relative energy perturbations to the $\delta p$ in this case. Similarly, it can also be checked that 
\begin{equation}
c_s^2=c_a^2=\frac{p^{(0)}{'}}{\rho^{(0)}{'}},
\label{cs2=ca2_mixto}
\end{equation}
after expanding the $\rho^{(0)}_i{'}$ in \eqref{cs2_mixto}. On the other hand, since $\delta q_1=0$ and $\delta q_2\propto\delta\phi_2$, and the l.h.s. of \eqref{delta p gen 2kin} does not depend on $\delta\phi_i$. Thus, since the perturbation coefficients are gauge invariant and $\delta q_r$ is generally non-zero, we have that
\begin{equation}
\sigma=0, 
\label{sigma_0_mixto}
\end{equation}
 Thus, there is also no contribution from the relative momentum perturbation in this case and the only non-adiabatic contribution to the pressure perturbation is due to the relative energy density perturbations.

\subsubsection{Power-law coupling functions}
We will now further study the case in which the coupling functions are power-laws of $Y$, i.e., $H_i(Y)\propto\lambda_i Y^{\alpha_i}$. The speed of sound will thus read
\begin{equation}
    c_s^2=\frac{w_1\,\rho_1^{(0)}{'}+c_2^2\,\rho_2^{(0)}{'}}{\rho_1^{(0)}{'}+\rho_2^{(0)}{'}},
    \label{cs2_mixto_step1_cute}
\end{equation}
where $w_1=-1$  and $w_2$ is given by \eqref{w_asympt_2kin}. This can be rewritten as follows when using the conservation equations \eqref{cons1Q} and \eqref{cons2Q}:
\begin{equation}
    c_s^2=\frac{(\rho_1^{(0)}+p_1^{(0)})\,w_1+(\rho_2^{(0)}+p_2^{(0)})\,c_2^2}{\rho^{(0)}+p^{(0)}}-\frac{Q\,(c_2^2-w_1)}{(\rho_1^{(0)}{'}+\rho_2^{(0)}{'})},
    \label{cs2_mixto_Q}
\end{equation}
where we can appreciate clearly the \textit{free} contribution and the contribution associated with the interaction, proportional to $Q$, which will only vanish if the kinetic field has $c_2^2=w_1=-1$, which is not possible  as it can be seen from equation \eqref{w_asympt_2kin}. For the $\kappa$ coefficient, we obtain that 
\begin{equation}
    \kappa=-2\frac{\rho_1^{(0)}{'}\rho_2^{(0)}{'}}{(1+\alpha_2)(\rho_1^{(0)}{'}+\rho_2^{(0)}{'})\,\rho^{(0)}},
    \label{kappa_mixto_Q}
\end{equation}
which reads the following way when recalling the conservation equations: 
\begin{align}
    \kappa&=\frac{2\rho{^{(0)}}^{-1}}{(1+\alpha_2)3\mathcal{H}(\rho^{(0)}+p^{(0)})}\left[\frac{9}{a^2}(\rho_1^{(0)}+p_1^{(0)})(\rho_2^{(0)}+p_2^{(0)})\right. \nonumber \\ 
    &\left. +\frac{3Q}{a^2\mathcal{H}}(\rho_1^{(0)}+p_1^{(0)})-\frac{3Q}{a^2\mathcal{H}}(\rho_2^{(0)}+p_2^{(0)})-\frac{Q^2}{a^2\mathcal{H}^2}\right].
    \label{kappa_Q_mixto_final}
\end{align}
Notice once again that $\kappa$ will generally be non-zero unless we are in the single-field domination regimes.

We will now focus on discussing the different scenarios for these models and study their stability.  As we have summarized in section 3, if $\alpha_1<0$, the kinetic field will be dominant at early times and $Q>0$, allowing the asymptotic $\phi_1$ domination regime to be reached. On the other hand, if $0<\alpha_1<1$, the kinetic field will dominate at early  times but $Q<0$ and the asymptotic $\phi_1$ domination regime is not reached, which results in both fields tracking each other in the future. We will discuss both cases separately, and with this in mind we will express $c_s^2$ and $\kappa$ in a more physical way. Therefore, we will introduce the density parameters of each field: 
\begin{align}
    \Omega_1=&\frac{\rho_{1,0}^{(0)}}{\rho_\text{crit}}=\frac{C_1}{\rho_\text{crit}}(1-\alpha_1),
    \label{Omega1_mixto}\\
    \Omega_2=&\frac{\rho_{2,0}^{(0)}}{\rho_\text{crit}}=\frac{C_2}{\lambda_2\rho_\text{crit}}(1+\alpha_2),
    \label{Omega2_mixto}
\end{align}
where the zero sub-index denotes the current time ($a=1$). Deriving the constraint \eqref{constrain_mixto} with respect to $a$ we can write
\begin{equation}
    Y'(a)=-\frac{6C_2\alpha_2Y}{a^7C_1Y^{\alpha_1+\alpha_2}(\alpha_1-1)\alpha_1+aC_2\alpha_2(1+\alpha_2)},
    \label{Yprima_mixto}
\end{equation}
which, when introduced in \eqref{cs2_mixto} allows us to finally obtain, after simplifications,
\begin{equation}
    c_s^2=-\frac{a^6\Omega_1Y^{\alpha_1+\alpha_2}\alpha_1+\Omega_2\alpha_2\frac{\alpha_2-1}{1+\alpha_2}}{\alpha_2\Omega_2-a^6\Omega_1Y^{\alpha_1+\alpha_2}\alpha_1},
    \label{cs2_mixto_physical}
\end{equation}
and
\begin{align}
    \kappa&=-12\frac{\Omega_1\Omega_2Y^{\alpha_1}\alpha_1\alpha_2(\Omega_1Y^{\alpha_1}+\Omega_2Y^{-\alpha_2}a^{-6})^{-1}}{a(1+\alpha_2)^2[a^6Y^{\alpha_1+\alpha_2}\alpha_1\Omega_1-\alpha_2\Omega_2]^2}\nonumber \\ 
    &\times[a^6Y^{\alpha_1+\alpha_2}\alpha_1(1+\alpha_2)\Omega_1-\alpha_2\Omega_2].
    \label{kappa_mixto_physical}
\end{align}

\begin{figure}[h]
     \centering
     \includegraphics[scale=0.58]{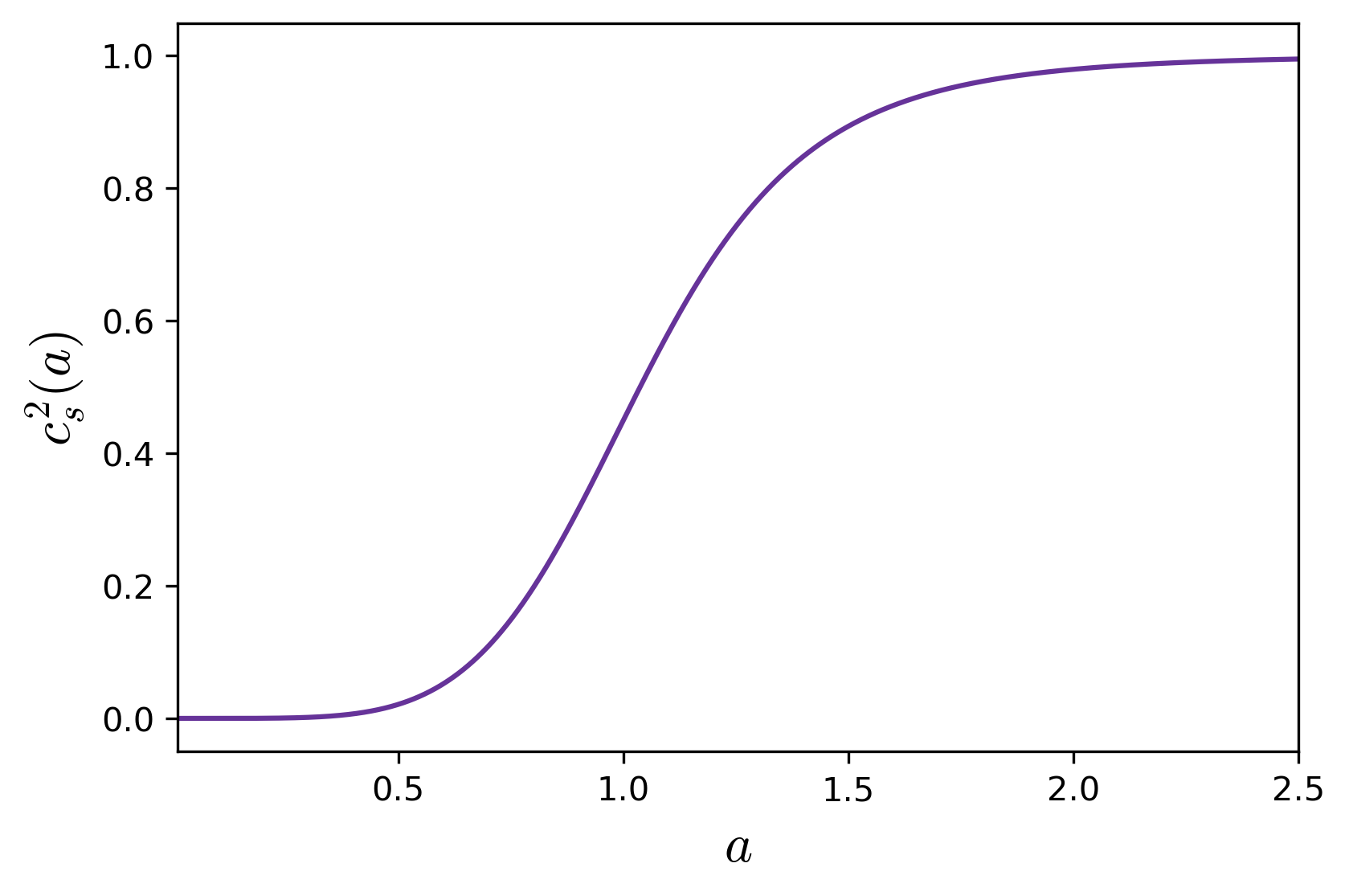}
     \caption{$c_s^2$ in terms of $a$ for $\alpha_1=-7/10$ and $\alpha_2=1$.}
     \label{Fig3}
\end{figure}
\begin{figure}[h]
    \centering
     \includegraphics[scale=0.55]{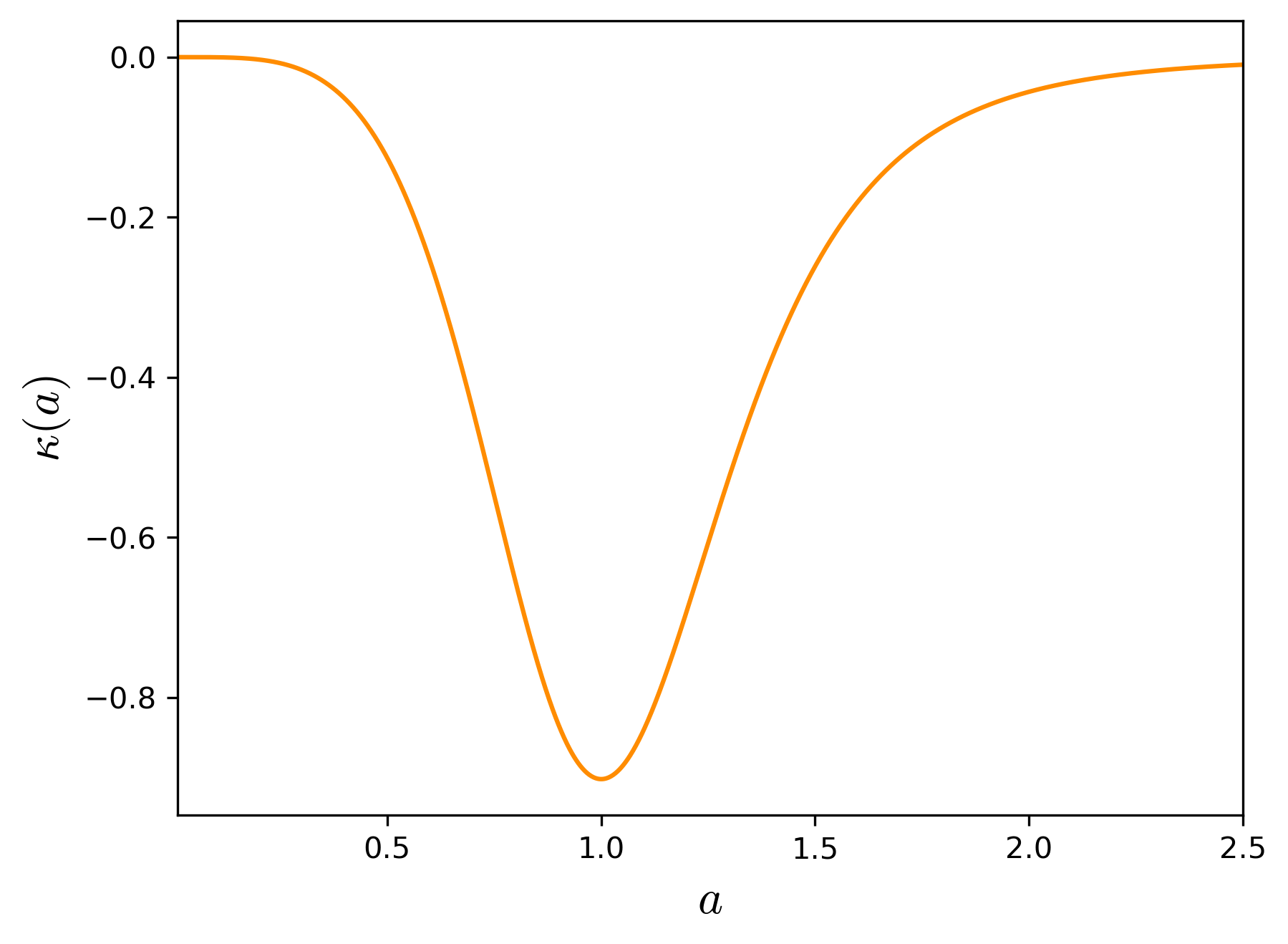}
     \caption{${\kappa}$ in terms of $a$ for $\alpha_1=-7/10$ and $\alpha_2=1$.}
     \label{Fig4}
\end{figure}

In Fig.\ref{Fig3} and Fig.\ref{Fig4} we have represented both $c_s^2$ and $\kappa$ for certain values of the parameters $\alpha_1,$ $\alpha_2$ and $C_1$, for the case in which the $\phi_1$ field behave as phantom energy at early times ($\alpha_1<0$).
As we can appreciate in these figures, $c_s^2$ will transition from $c_s^2=w_2=0$ during the $\phi_2$ domination epoch to $c_s^2=1$ behavior under $\phi_1$ domination. This happens because, even if the $\phi_1$ domination regime is reached for $\alpha_1<0$, the potential field will not be an exact cosmological constant and it has its own perturbations, while the kinetic field will always present stiff behavior ($w_\mathrm{eff,2}\simeq 1$) under potential domination as a consequence of the interaction. In fact, this is physically illustrated through \eqref{cs2_mixto_physical}, where it can be checked that it is the same expression  as $w_\mathrm{eff,2}(a)$ given by \eqref{weff2mixto}. On the other hand, the $\kappa$ coefficient will tend to zero in the single-field domination regimes for $\alpha_1<0$, but it will become relevant in the intermediate regime as a result of the interactions taking place between the fields. Furthermore, 
as in the shift-symmetric case, $\delta p\simeq c_a^2\delta\rho$ in the asymptotic regimes and the evolution will approximately be adiabatic. However, in this case $c_s^2$ will always be positive if $w_2>0$ and the models will thus be stable, which is very interesting from a phenomenological point of view. When it comes to the particular dark matter-dark energy interactions ($\alpha_2=1$), $c_s^2$ becomes much bigger than zero as $\phi_1$ becomes dominant, which could be problematic for the matter power spectrum. 

\begin{figure}[h]
     \centering
     \includegraphics[scale=0.57]{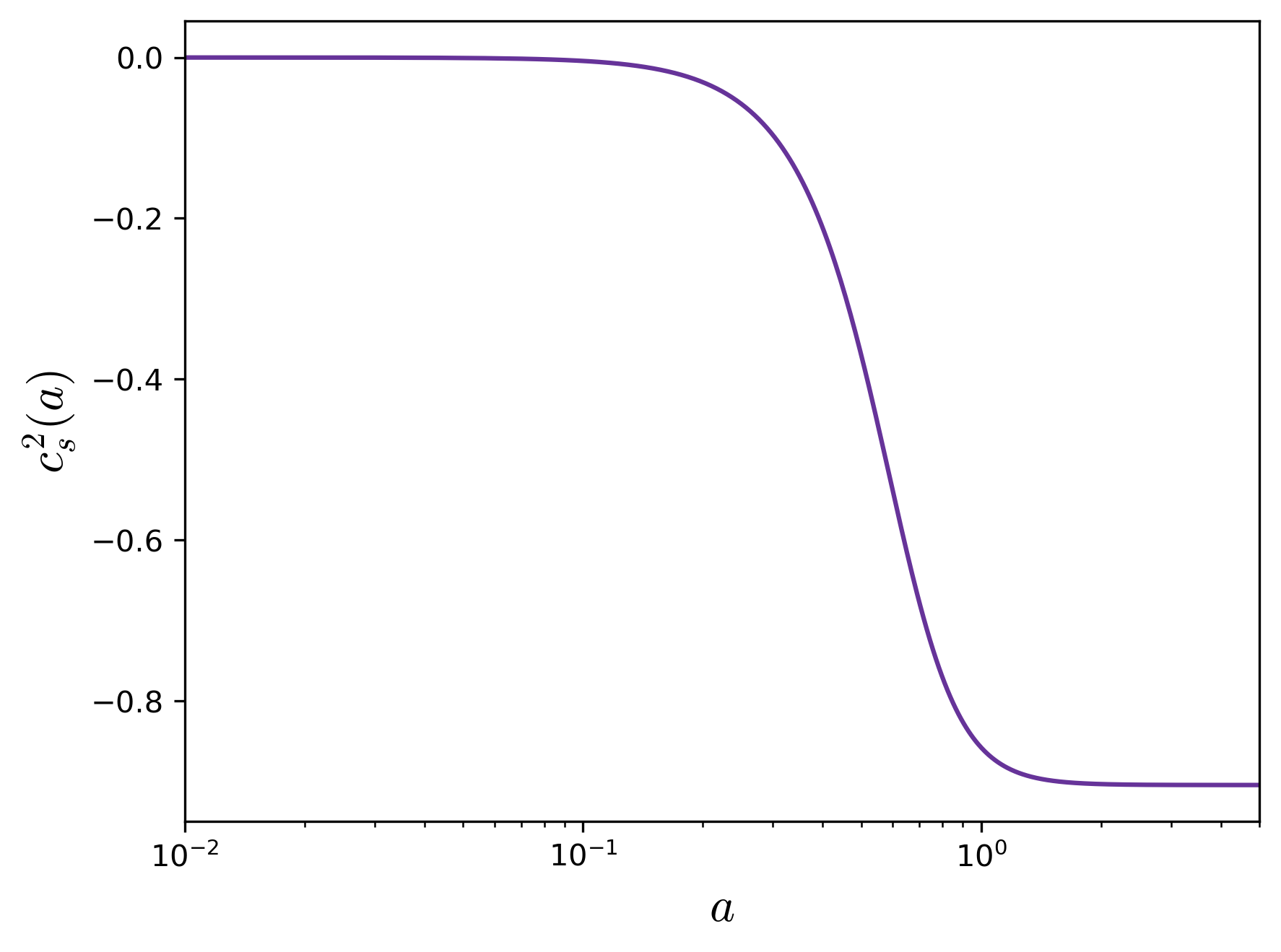}
     \caption{$c_s^2$ in terms of $a$ for $\alpha_1=1/20$ and $\alpha_2=1$ (logarithmic scale on $a$ axis).}
     \label{Fig5}
\end{figure}

\begin{figure}[h]
    \centering
     \includegraphics[scale=0.57]{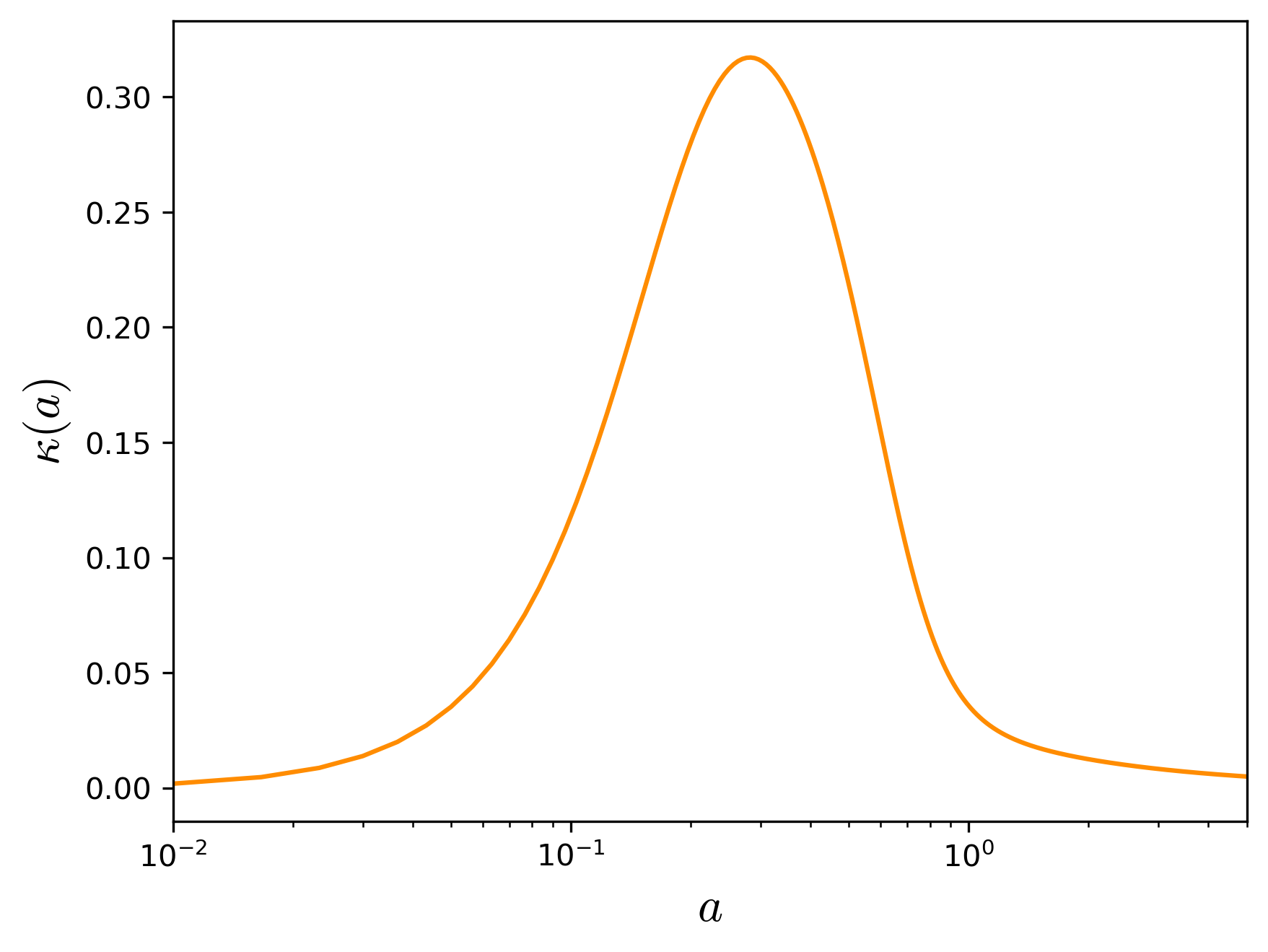}
     \caption{$\kappa$ in terms of $a$ for $\alpha_1=1/20$ and $\alpha_2=1$ (logarithmic scale on $a$ axis).}          
     \label{Fig6}
\end{figure}

In Fig.\ref{Fig5} and Fig.\ref{Fig6} both $c_s^2$ and $\kappa$ are represented for certain values of the parameters $\alpha_1$, $\alpha_2$ and $C_1$, for the tracking case ($0<\alpha_1<1$). 
As we can see, $c_s^2$ will transition from  $c_s^2=w_2=0$ during the single-field $\phi_2$ domination regime to $w_\mathrm{track}=-0.860$ given by \eqref{w_tracking}. On the other hand, $\kappa$ tends to zero both at early (when the kinetic field is dominant) and late times as the tracking regime is reached. This fact,  can be analytically checked by recalling the asymptotic behaviour of $Y(a)$ for $a\rightarrow\infty$ in this case ($Y(a)\propto a^{-\frac{6}{\alpha_1+\alpha_2}}$, see \eqref{constr_asymptotic_mixto}). As a consequence, $\delta p\simeq c_a^2\delta \rho$ in that regime and the evolution will be asymptotically adiabatic, with both components presenting the same effective behavior. Lastly,  as $c_s^2>0$ for $w_2>0$ and $w_\mathrm{track}>0$,  one can expect those models to be stable and, therefore, of special interest. 

\subsection{Diff vs TDiff comparison}\label{sec4p3}
Lastly, in this subsection we will discuss the main aspects of the TDiff models discussed in previous sections and compare them with the analogous cases in Diff theories. We will classify the models into two categories: single- and multi-field models; and discuss the different scenarios when it comes to the adiabaticity and the different contributions to the pressure perturbations. 

\subsubsection{Single-field models}
On the one hand, general single-field models with a kinetic and a potential contribution are not adiabatic in general ($c_s^2\neq c_a^2$), both in the Diff and in the TDiff case \footnote{Nevertheless, there can be particular cases in which perturbations are adiabatic, like Diff fields than can be described as barotropic perfect fluids.}. On the other hand, if the fields are kinetically driven, the pressure perturbations will be adiabatic ($c_s^2=c_a^2$), both in the Diff and TDiff scenarios. However, while in the Diff case the only possible phenomenology for a kinetically dominated field is that of a stiff fluid ($w=c_s^2=1$), in TDiff models we can reproduce a wide range of possible phenomenological behavior depending on the coupling function $H_K(Y)$.

\subsubsection{Multi-field models}
On the one hand, as we have discussed, TDiff multi-field models will be non-adiabatic in general and present relative energy and momentum contributions to the pressure perturbations  ($c_s^2\neq c_a^2$, $\kappa\neq0$, $\sigma\neq 0$). A similar reasoning can be applied to non-interacting multi-field Diff theories, allowing us to infer the same conclusion for them. 

When it comes to the particular cases we studied  in deeper detail, we first have the shift-symmetric multi-field TDiff model. In this case, $\sigma=0$ and $c_s^2=c_a^2$ (regardless of the coupling functions $H_{i}(Y)$), but perturbations are not adiabatic since $\kappa\neq0$, even if the speed of sound is equal to the adiabatic speed of sound. The analogous Diff model would involve two free stiff fluids, which in reality can be interpreted as a single stiff fluid. Thus, this model would be adiabatic with $c_s^2=c_a^2=1$ and $\kappa=0$, $\sigma=0$. As we can see, the freedom to have different coupling functions for each field in the TDiff case allows us to have different types of components (which interact as a consequence of the Diff symmetry breaking) and to present non-adiabatic perturbations as a consequence of there being relative energy perturbations, even if $c_s^2=c_a^2$.

The other particular model of interest we studied is the mixed-regime multi-field TDiff model, for which we obtained that $\sigma=0$ and $c_s^2=c_a^2$ (again, regardless of the coupling functions $H_i(Y)$), but $\kappa\neq0$ and perturbations are not adiabatic. The analogous Diff model would involve a stiff field and a cosmological constant-like component with no interactions between them. Thus, we would obtain that $c_s^2=c_a^2=1$ and $\kappa=\sigma=0$, and the model would therefore be adiabatic\footnote{It is worth mentioning that we are neglecting $p'_1$ with respect to $p'_2$ and $\rho'_1$ with respect to $\rho'_2$ in the calculation of $c_a^2$.} \cite{Unnikrishnan:2024agf}. Once again, it is the Diff symmetry breaking (which allows us to have freedom in the coupling functions $H_i(Y)$) which affects the possible phenomenology and introduces the non-adiabatic perturbations with respect to the analogous Diff case.

As an interesting comment, it is worth noting that $c_s^2=c_a^2$ and $\sigma=0$ in the TDiff shift-symmetric and mixed regime two-field models. This is due to the fact that in both cases there is only one independent perturbation for each field. Therefore, from the point of view of perturbations, those TDiff models are similar to multi-field Diff models involving two non-interacting fields that can be described as perfect fluids \cite{Unnikrishnan:2024agf}, since they have $c_s^2=c_a^2$, $\kappa\neq0$. Lastly, we want to emphasize that multi-field TDiff models are necessarily interacting, whereas that is not necessarily the case for Diff fields.
A summary of the models discussed in this subsection is included in Table \ref{Tab1}.

\begin{table*}[h]
\centering
\begin{tabular}{|c|c|c|}
\hline
                   & Diff                                                & TDiff                                                        \\ \hline
Single field (general)  & $c_s^2\neq c_a^2$, non-adiabatic                    & $c_s^2\neq c_a^2$, non-adiabatic                             \\ \hline
Single field  (kinetic)  & $c_s^2=c_a^2=1$, stiff adiabatic                    & $c_s^2=c_a^2=w(a)$, adiabatic                                \\ \hline
Two fields (kinetic) & $c_s^2=c_a^2=1$, $\kappa,\sigma=0$, stiff adiabatic & $c_s^2=c_a^2=f(a)$, $\kappa\neq0$, $\sigma=0$, non-adiabatic \\ \hline
Two fields (mixed)   & $c_s^2=c_a^2=1$, $\kappa,\sigma=0$, stiff adiabatic & $c_s^2=c_a^2=f(a)$, $\kappa\neq0$, $\sigma=0$, non-adiabatic \\ \hline
\end{tabular}
\caption{Summary of the families of models discussed throughout this section.}
\label{Tab1}
\end{table*}

\section{Some stable cases of interest}\label{sec5}
In this section, we will briefly discuss some specific stable cases with $c_s^2\geq0$ which could be interesting in different cosmological scenarios.

\subsection{Shift-symmetric with $w_i>0$}
 We have introduced the shift-symmetric TDiff two-field models at the background and at the perturbative level in sections \ref{sec221} and \ref{sec4p1}, respectively. In particular, we have considered in detail shift-symmetric models with power-law coupling functions. For that case, cosmological models with positive EoS parameters, that is $w_i>0$, will be stable. In this scenario, the energy exchange will always favor the field with the lower EoS parameter \cite{Tessainer1} and the single-field domination regimes will be reached, as we saw in Section \ref{sec2}. Thus, as discussed in the previous section, the speed of sound $c_s^2$ will transition from one $w_i$ to the other (from the one with a larger value of $w_i$ to that with a smaller one), presenting a peak in the intermediate regime, when the interaction is stronger. Consequently, $c_s^2$ will be positive if both EoS parameters are positive and there will be a family of stable models. 

\begin{figure}[h]
    \centering
     \includegraphics[scale=0.55]{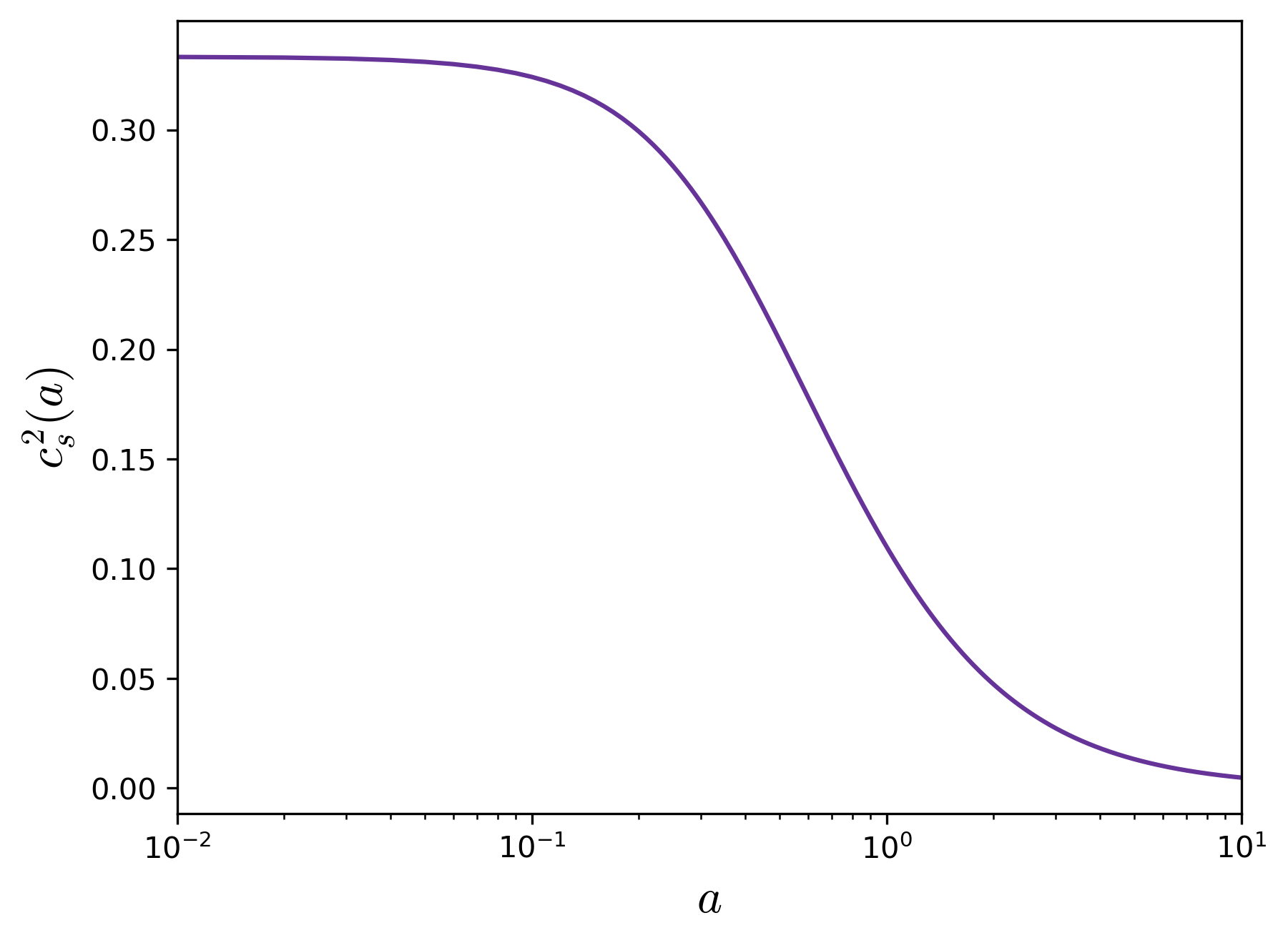}
     \caption{$c_s^2$ in terms of $a$ for $\alpha_1=1/2$ (dark radiation) and $\alpha_2=1$ (dark matter).}
     \label{fig2kinstable}
\end{figure}

For instance, a particular case of physical interest corresponds to $w_1=1/3$ (asymptotic dark radiation) and $w_2=0$ (asymptotic dark matter). In this scenario, $\phi_1$ would dominate in the past and $\phi_2$ would do so in the future, allowing us to describe dark radiation-dark matter interactions within this framework. The presence of dark radiation would alter the effective behavior of the dark sector at early times (see Fig.\ref{fig2kinstable} for the effective speed of sound).

\subsection{Shift-symmetric dark sector}
\begin{figure}[h]
    \centering
     \includegraphics[scale=0.57]{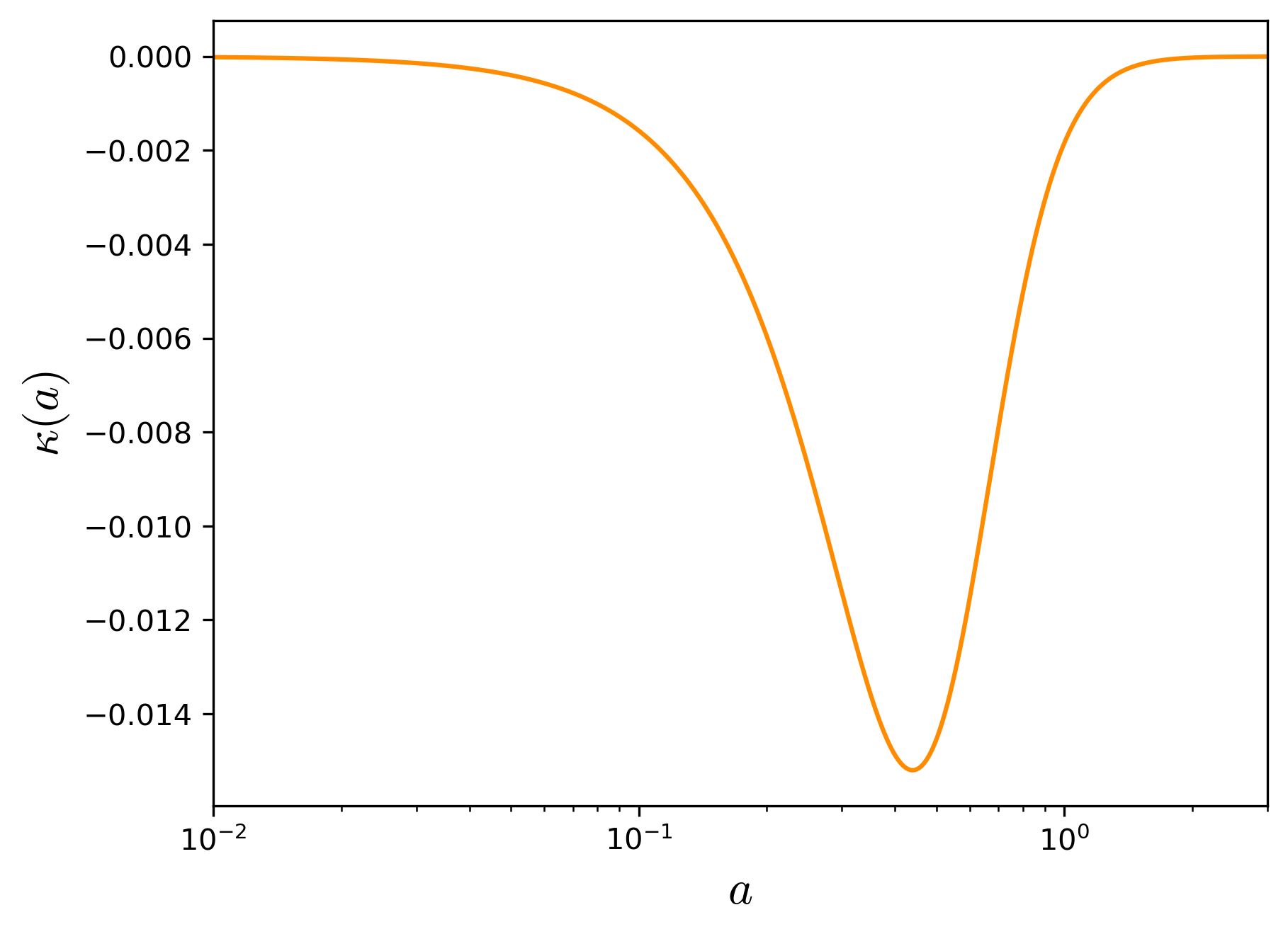}
     \caption{$\kappa$ in terms of $a$ for $C_1=0.30$, $C_2=0.70$, $\beta_1=-2.00$, $\beta_2=-4.67$ in adequate units, such that $\Omega=\Omega_c/\Omega_\Lambda=0.36$.}          
     \label{Fig8}
\end{figure}

Let us focus again on shift-symmetric TDiff two-field models (see sections \ref{sec221} and \ref{sec4p1}). As we have already discussed, if the coupling functions are power-laws, dark sector models eventually become unstable at late times (when the fluid with $w_\mathrm{DE}<-1/3$ dominates) and, moreover, $c_s^2$ becomes considerably larger than zero at some point during the intermediate regime thus spoiling structure formation.
However, for certain shapes of the coupling functions $H_i(Y)$ it is in fact possible to have $c_s^2=0$ identically. More concretely, if we fix one of the coupling functions, e.g., $H_1(Y)$, $c_s^2$ in \eqref{cs_2kin_gen} will be equal to zero if the other ($H_2(Y)$) is such that
\begin{align}
    H_2''&=-\frac{C_1}{H_1^2C_2(H_2C_1+H_1C_2)}(H_1''H_2^3C_1-2H_1'^2H_2^2C_2\nonumber\\ &+H_1H_1''H_2^2C_2+4H_1H_1'H_2H_2'C_2-2H_1^2H_2'^2C_2)
    \label{ED_H2}
\end{align}
is satisfied. This ODE has an analytical solution if $H_1(Y)=\lambda Y$, where $\lambda$ is a constant. In this particular case, we will have that 
\begin{equation}
    H_2(Y)=-\frac{Y(Y+C_1C_2\lambda\beta_1+YC_1C_2\lambda\beta_2)}{C_1^2(\beta_1+Y\beta_2)},
    \label{H_2_analytical}
\end{equation}
where $\beta_i$ are integration constants. This allows us to obtain $Y(a)$ analytically from the constraint \eqref{constraint_2kin_gen} as its positive solution: 
\begin{align}
    Y(a)&=\frac{1}{a^3(C_g+2C_1C_2C_g\beta_2+C_1^2C_2^2C_g\beta_2^2)}\left(-a^3(C_1C_2 C_g\beta_1\right. \nonumber \\ 
    &\left. -C_1^2C_2^2C_g\beta_1\beta_2)\right. &  \nonumber \\ 
    &\left. +\sqrt{C_1 C_g+3C_1^2 C_2 C_g\beta_2+3C_1^3 C_2^2 C_g\beta _2^2+C_1^4 C_2^3 C_g\beta_2^3}\right),
    \label{Y(a)_analytical_LCDM}
\end{align}
where the $\lambda$ factors were reabsorbed into the $C_i$ constants. Notice that this results in both energy densities presenting a dynamical scaling behavior. With respect to these solutions, on the one hand, it is worth noting from \eqref{Y(a)_analytical_LCDM} that $\beta_2>-1/(C_1C_2)$ in order to avoid complex $Y$ values. On the other hand, $H_2(Y)$ must also be positive, which imposes some restrictions in the integration constants as well. More specifically, as we can appreciate from \eqref{H_2_analytical}, both $\beta_1$ and $\beta_2$ must be negative and 
\begin{equation}
    Y(a)>-\frac{C_1C_2\beta_1}{1+C_1C_2\beta_2}
    \label{Y(a)_foralla_LCDM}
\end{equation}
must be satisfied for all $a$. Taking this into account, substituting \eqref{Y(a)_analytical_LCDM} in \eqref{rho_i_shift_symmetric} and computing the EoS parameter of the model yields:
\begin{equation}
    w=\frac{p_1+p_2}{\rho_1+\rho_2}=\frac{-1}{1+\Omega a^{-3}},
    \label{w_LCDM}
\end{equation}
where 
\begin{equation}
    \Omega\equiv-\frac{2(1+C_1C_2\beta_2)^2}{C_2\beta_1\sqrt{C_1C_g(1+C_1C_2\beta_2)^3}}.
    \label{f_constant}
\end{equation}
which is identical to total dark sector equation of state of
$\Lambda$CDM with $\Omega$ playing the role of the ratio 
$\Omega_c/\Omega_\Lambda$ of $\Lambda$CDM. Thus, this type of models behave at the background level as $\Lambda$CDM, their perturbations are stable with $c_s^2=0$, but however the total pressure perturbation is non-vanishing since $\kappa\neq 0$ (see Fig.\ref{Fig8}).

\subsection{Mixed-regime tracking matter}
\begin{figure}[h]
    \centering
     \includegraphics[scale=0.55]{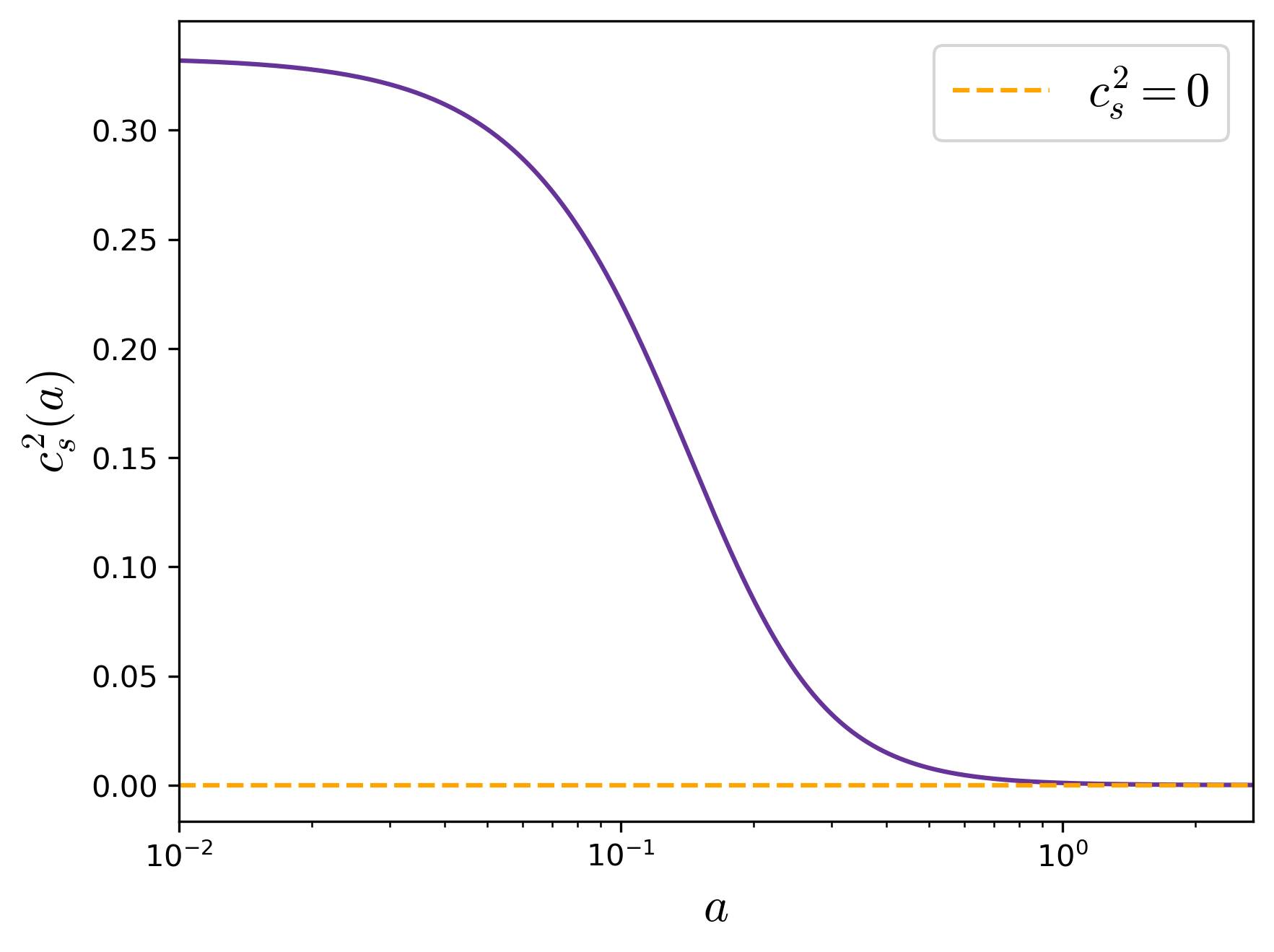}
     \caption{Effective speed of sound $c_s^2$ in terms of the scale factor.}
     \label{weff_total_tracking}
\end{figure}

Lastly, let us focus on an interesting model of the mixed-regime case, presented in sections \ref{sec222} and \ref{sec4p2} at the background and perturbation level, respectively. As we discussed, if $0<\alpha_1<1$, the single potential domination regime is not reached and both the kinetic and the potential fields end up tracking each other in the future, with an equation of state parameter given by (see \eqref{w_tracking})
\begin{equation}
    w_\mathrm{track}=\frac{\alpha_1-\alpha_2}{\alpha_1+\alpha_2}.
    \label{w_tracking2}
\end{equation}
  In this case, $c_s^2$ transitions from $w_2$ to $w_\mathrm{track}$. Thus, if $\alpha_1=\alpha_2$, then both components will track each other with matter behavior $w_\mathrm{track}=0$ and $c_s^2$ will tend to zero. This opens up a wide range of possible phenomenological scenarios, since we could describe an effective dark matter behavior with $c_s^2=0$ as a result of TDiff field interactions between a kinetic field with an asymptotic behavior given by $w_2=(1-\alpha_2)/(1+\alpha_2)$ and a potentially driven field with $w_1=-1$, both coupled to gravity through the same power-law functions. Indeed, $\phi_2$ will asymptotically scale as expected from $w_2$ in the past, and $\phi_1$ will behave the following way in such regime (see \eqref{w_eff_asymp_1_kindom_powerlaw}):
\begin{equation}
    w_\mathrm{eff,1}=\frac{\alpha_2-1}{1+\alpha_2}=-w_2,
    \label{weff_1_analytical_kindom}
\end{equation}
with $-1<w_\mathrm{eff,1}<0$. For instance, we could describe dark matter as a consequence of the effective interaction of two TDiff fields, a kinetic one with $\alpha_2=1/2$ (dark radiation, $w_2=1/3$) and a potential one with $\alpha_1=1/2$. So, this is a dark radiation-dark matter stable model, which  will introduce a short \textit{warm} epoch as a result of the interactions (see Fig.\ref{weff_total_tracking}).

\section{Conclusions}\label{conclusions}
In this work, we have developed the theory of cosmological perturbations of scalar fields models breaking Diff invariance down to TDiff invariance through the matter sector in models involving one or  multiple scalar fields. We have used the covariantized approach, in which the theory is rewritten in a Diff invariant way through the introduction of an extra Stueckelberg-like vector field whose EoM is equivalent to the constraint of the TDiff formalism. In the multi-field case, this constraint presents contributions from the different fields and induces a natural mechanism for effective interactions between the fields.

With this in mind, we developed the cosmological perturbation formalism for TDiff invariant theories in the matter sector and computed the contributions to the pressure perturbation and the speed of sound . Firstly, we considered the single-field scenario, discussing the general case, as well as the kinetic and potential domination regimes. Although in the general case perturbations are non-adiabatic, in the particular case in which the kinetic term of the field dominates over the potential one, the adiabatic condition ($c_s^2=c_a^2$) is satisfied. 

Afterwards, we focused on the study of cosmological perturbations in multi-field case, discussing in detail two particular models: the shift-symmetric model (two kinetically driven fields) and the mixed-regime model (a kinetically driven field and a field dominated by its potential term). We first performed a general analysis of the pressure perturbation, showing that it has four independent contributions that depend on the total density perturbation, the total momentum perturbation, and the relative density and momentum perturbations. However, in the particular models studied in this work, only the first three are present due to having only one independent perturbation per field. Similarly, we obtained that these models will always have $c_s^2=c_a^2$, but they will not be adiabatic in general since there will be a contribution proportional to the relative energy perturbation, with a coefficient $\kappa\neq 0$, so that $\delta p\neq c_a^2\delta\rho$. We also compared TDiff models with the analog Diff models, seeing that the multi-field interacting models in the particular regimes we investigated are comparable to Diff models with non-interacting fluids at the perturbation level.

On the one hand, we analyzed the shift-symmetric case and studied the particular case of power-law coupling functions in detail. In this case, we have seen that, since the single-field domination regimes are reached, the effective speed of sound transitions from one EoS parameter value $w_i$ to the other, presenting a peak in the intermediate regime in which the interaction becomes stronger. The $\kappa$ coefficient is approximately zero in the single-field domination regimes, but becomes large in magnitude in the intermediate regime as a consequence of the interactions. In terms of the stability, we have seen in light of this analysis that models where $w_{1,2}>0$ will be stable ($c_s^2>0$). 

We have also studied the mixed-regime case, focusing on the specific case of power-law coupling functions. Within this model, there are two possible scenarios. If $\alpha_1<0$, the potential field gains energy from the kinetic one reaching its asymptotic domination regime in the future and thus behaving as a cosmological constant. $c_s^2$ smoothly transitions from $w_2$ (the EoS parameter of the kinetic field) to one, which is the effective behavior for the kinetic component under the potential-field domination regime. When it comes to $\kappa$, it tends to zero in the asymptotic regimes but becomes larger in the intermediate regimes as a consequence of the interactions. However, if $0<\alpha_1<1$, the potential field loses energy in favor of the kinetic one and will not be able to reach its asymptotic domination regime, which results in both components tracking each other in the future. Regarding the speed of sound, it will smoothly transition from $w_2$ to the tracking value of the effective EoS parameter ($w_\mathrm{track}$), and $\kappa$ will tend to zero except in the intermediate regimes, when the interaction becomes stronger. When it comes to stability in terms of the speed of sound, models with positive $w_2$ (and $w_\mathrm{track}>0$ if $0<\alpha_1<1$) will be stable.

Lastly, we also discussed briefly some possible cases of interest among models with $c_s^2>0$. These models include cases such as shift-symmetric models with $w_i>0$, which could be useful to describe dark radiation-dark matter interactions. There is also the possibility to describe dark matter-dark energy interacting models with $c_s^2=0$ within the context of shift-symmetric models if the coupling functions of each field are related in a certain  way, which would result in an effective EoS evolution identical to $\Lambda$CDM, but with $\kappa\neq 0$. In addition, in the mixed-regime case it is possible to have a dark matter type of tracking if the kinetic and potential fields have the same power-law coupling function.

Future projects include a deeper study of a broader set of models, both at the background and at the perturbative level, as well as investigating the particular applications of the models of interest presented in this work regarding the dark sector. Furthermore, a full observational analysis could be performed in order to compare the theory with observational data. 

\section*{Acknowledgements}
The authors would like to thank Darío Jaramillo-Garrido and Miguel Orbaneja-Pérez for useful comments and suggestions. DTB
acknowledges financial support from Universidad Complutense de Madrid and Banco Santander through the Grant No. CT25-24. This work has been supported by the MICIN (Spain) Project No. PID2022-138263NB-I00 funded by MICIU/AEI/10.13039/501100011033 and by ERDF/EU.

\section{Appendix}\label{Appendix}
\subsection*{Perturbation coefficients in the shift-symmetric case}
The coefficients of the energy and pressure perturbations in the multi-field shift-symmetric case read
\begin{align}
    \mathcal{A}&=\frac{X_1H_1H_1''+X_2H_2''H_1-YX_2H_2''H_1'}{X_1H_1''+X_2H_2''},
    \label{A_2kin}\\
    \mathcal{B}&=\frac{X_1YH_1''H_2'}{X_1H_1''+X_2H_2''},
    \label{B_2kin}\\
    \mathcal{C}&=\frac{X_2YH_2''H_1'}{X_1H_1''+X_2H_2''},
    \label{C_2kin}\\
    \mathcal{D}&=\frac{X_1H_1''H_2+X_2H_2H_2''-YX_1H_1''H_2'}{X_1H_1''+X_2H_2''},   
    \label{D_2kin}\\
    \mathcal{E}&=\frac{X_1H_1''H_1+X_2H_2''H_1+YH_1'X_2H_2''-2X_1H_1'^2}{X_1H_1''+X_2H_2''},
    \label{E_2kin}\\
    \mathcal{F}&=\frac{-2X_1H_1'H_2'-X_1YH_1''H_2'}{X_1H_1''+X_2H_2''},
    \label{F_2kin}\\
    \mathcal{G}&=\frac{-2X_2H_1'H_2'-X_2YH_2''H_1'}{X_1H_1''+X_2H_2''},
    \label{G_2kin}\\
    \mathcal{J}&=\frac{X_2H_2''H_2+X_1H_1''H_2+YH_2'X_1H_1''-2X_2H_2'^2}{X_1H_1''+X_2H_2''}.
    \label{J_2kin}
\end{align}
\subsection*{Perturbation coefficients in the mixed-regime case}
The coefficients of the energy and pressure perturbations in the multi-field mixed-regime case read
\begin{align}
    \hat{\mathcal{A}}&=-V'(H_1-YH_1')+YH_1''V\frac{H_1'V'}{H_2''X_2-H_1''V},
    \label{A_2kin_mixto}\\
    \hat{\mathcal{B}}&=-\frac{YH_1''VH_2'}{H_2''X_2-H_1''V},
    \label{B_2kin_mixto}\\
    \hat{\mathcal{C}}&=-\frac{H_1'V'X_2YH_2''}{H_2''X_2-H_1''V},
    \label{C_2kin_mixto}\\
    \hat{\mathcal{D}}&=H_2-YH_2'+\frac{H_2'X_2YH_2''}{H_2''X_2-H_1''V},   
    \label{D_2kin_mixto}\\
    \hat{\mathcal{E}}&=-\hat{\mathcal{A}},
    \label{E_2kin_mixto}\\
    \hat{\mathcal{F}}&=-\hat{\mathcal{B}},
    \label{F_2kin_mixto}\\
    \hat{\mathcal{G}}&=\frac{X_2 H_1' V'}{H_2''X_2-H_1''V}(2H_2'+YH_2''),
    \label{G_2kin_mixto}\\
    \hat{\mathcal{J}}&=H_2+YH_2'-\frac{X_2H_2'}{H_2''X_2-H_1''V}(2H_2'+YH_2'').
    \label{J_2kin_mixto}
\end{align}

\newpage
\bibliographystyle{elsarticle-num} 
\bibliography{bibliografia} 

\end{document}